\documentclass[pdflatex,sn-mathphys-num,iicol]{sn-jnl}
\usepackage[utf8]{inputenc}
\usepackage{cuted}
\usepackage{lineno}
\usepackage{graphicx}%
\usepackage{latexsym}%
\usepackage{amsmath,amssymb,amsfonts}%
\usepackage{amsthm}%
\usepackage{mathrsfs}%
\usepackage[title]{appendix}%
\usepackage{textcomp}%
\usepackage{manyfoot}%
\usepackage{booktabs}%
\usepackage{algorithm}%
\usepackage{algorithmicx}%
\usepackage{algpseudocode}%
\usepackage{listings}%
\theoremstyle{thmstyleone}%
\usepackage{todonotes}
\theoremstyle{thmstyletwo}%
\theoremstyle{thmstylethree}%
\usepackage{textcomp}
\usepackage{rotating}
\usepackage{bm}
\usepackage{float}
\usepackage{balance}
\usepackage{etoolbox}
\PassOptionsToPackage{colorlinks=true,linkcolor=blue,citecolor=blue, urlcolor=blue,anchorcolor = blue}{hyperref}
\usepackage[colorlinks=true,linkcolor=blue,citecolor=blue, urlcolor=blue,anchorcolor = blue]{hyperref}
\usepackage{epsf}
\usepackage{graphics}
\usepackage[export]{adjustbox}
\usepackage{tabularx}
\usepackage{multirow}
\usepackage{array} 
\usepackage{makecell}  
\usepackage{diagbox}  
\usepackage{colortbl}  
\usepackage{diagbox}
\usepackage{slashbox}
\usepackage{epsfig,latexsym}
\usepackage[section]{placeins}
\usepackage{booktabs}
\usepackage{changes}
\usepackage{array, makecell, booktabs, xcolor}

\def\bea{\begin{eqnarray}}
\def\eea{\end{eqnarray}}

\begin{document}


\title{Classifying Urban Regions by Aggregated Pollutant–Weather Correlation Strength: A Spatiotemporal Study} 

\author[1]{\fnm{Koyena} \sur{Ghosh}} 
\equalcont{These authors contributed equally to this work.}
\author*[2]{\fnm{Suchismita} \sur{Banerjee}} \email{suchib@bose.res.in}
\equalcont{These authors contributed equally to this work.}
\author[2]{\fnm{Urna} \sur{Basu}}
\author[3]{\fnm{Banasri} \sur{Basu}} 

\affil[1]{\orgname{Maulana Abul Kalam Azad University of Technology}, \orgaddress{\state{West Bengal}, \postcode{741249}, \country{India}}}

\affil[2]{\orgname{S. N. Bose National Centre for Basic Sciences}, \orgaddress{\city{Kolkata}, \postcode{700106}, \country{India}}}

\affil[3]{\orgname{Indian Statistical Institute}, \orgaddress{\city{Kolkata}, \postcode{700108}, \country{India}}}

\abstract{Understanding pollutant–meteorology interactions is essential for environmental risk assessment. This study develops an entropy-based statistical framework to analyze static and temporal dependencies between urban air pollutants and meteorological variables across multiple Indian cities. Dependence is quantified using complementary linear and nonlinear measures, including Pearson correlation, mutual information, and relative conditional entropy. A key methodological contribution is a PCA-based composite indexing framework  that integrates these heterogeneous metrics into a unified and interpretable correlation score. For each pollutant–meteorological pair within a city, PCA is used to extract a joint variability index, while spatial variability is assessed by aggregating correlations across cities. These indices are further combined to derive a comprehensive city-level correlation score that represents overall pollutant–meteorology coupling strength and enables classification of cities into distinct interaction regimes. Sensitivity analysis, performed by systematically excluding individual variable pairs, demonstrates the robustness of the framework, with no single pair exerting disproportionate influence. Temporal dependencies are examined using transfer entropy and time-delayed mutual information. Results indicate that relative humidity generally leads changes in pollutant concentrations, whereas ambient temperature tends to lag, highlighting contrasting causal influences. Mutual information peaks at zero lag and decays rapidly, indicating strong short-term interactions with limited persistence. Overall, the proposed framework provides a unified and interpretable approach for assessing complex pollutant–meteorology interactions across diverse locations  and time.}


\maketitle

\section{Introduction} \label{intro}
Air pollution remains one of the most pressing environmental health risks worldwide~\cite{WHO2021AQG}, particularly in urban regions where population density, industrial activity, and vehicular emissions converge to produce complex pollutant mixtures. 
Understanding how atmospheric conditions influence the formation, transport, and transformation of pollutants is key to characterize the underlying dynamics of air quality. Air pollution dynamics can be viewed as a classic example of an open, driven-dissipative system~\cite{Chaos_2010,Bak_2007,Nic_1977,Love_2015,Love_2013} where energy and matter continuously flow through chemical, physical, and weather-related processes, giving rise to persistent fluctuations in concentration levels.
These far-from-equilibrium dynamics generate complex spatiotemporal patterns~\cite{Run_2019, Sein_2016, Jac_1999,Wan_2024,Ant_2025}, shaped by both natural and anthropogenic factors and often described in terms of probability distributions rather than deterministic trends. From a physics perspective, such systems challenge classical equilibrium models and call for tools from nonlinear dynamics~\cite{Raga_1996}, stochastic processes~\cite{Hwa_2013}, and information theory~\cite{Gol_2022}.

Pollutants (such as PM\textsubscript{2.5}, PM\textsubscript{10}, SO\textsubscript{2} and NO\textsubscript{2} and many more) are not merely passive tracers; they undergo nonlinear transformations~\cite{Bin_2019}, interact through feedback mechanisms~\cite{Im_2022}, and are transported via advection-diffusion processes~\cite{Lin_2021}, all modulated  by meteorological (such as temperature,  precipitation etc.) forcing that introduce stochastic fluctuations.  Understanding  the interdependence between pollutants and weather variables is essential not only for accurate predictive modeling, environmental monitoring and air quality risk assessment, but also for uncovering emergent behavior in complex environmental systems~\cite{Cov_1999,San_2002}. 

A substantial body of literature~\cite{Row_2024,Nak_2025,Mal_2023} has examined the relationships between air pollutant concentrations and meteorological variables. However, these investigations~\cite{Bose_2023,Pawar_2023,Wang_2022} have been overwhelmingly dominated by the use of a {\it{single dependence measure}}, most commonly linear correlation metrics, and are therefore limited in their ability to capture the complex and potentially nonlinear interactions that govern pollutant–meteorology dynamics. 
Moreover, these  approaches  overlook both non-stationarity~\cite{Garsa_2023} and directional causality. To address these limitations, we adopt an entropy based information-theoretic approach to study the interdependence of pollutants and weather variables. Although a small number of studies have employed nonlinear dependence measures, including mutual information~\cite{gupta_2018}, these measures are typically applied in isolation and are not combined with linear metrics within a unified analytical framework. In parallel, multivariate techniques such as principal component analysis (PCA)  and clustering have largely focused on pollutant concentrations or meteorological variables independently~\cite{a,b}. However, the lack of a comprehensive measure integrating multiple forms of statistical dependence has constrained robust and comparable assessment of pollutant–meteorology interactions.

In this work, we address this issue 
by  adopting an entropy-based information-theoretic framework, rooted in statistical physics and complex systems science, which provides powerful tools to characterize structure and dynamics in multivariate systems with inherent fluctuations~\cite{Gol_2022,Andrea_2023}.
Metrics such as Shannon entropy~\cite{Shanon}, Mutual Information (MI)~\cite{Fras_1986,Steu_2002}, and Transfer Entropy (TE)~\cite{TE_2000} have proven valuable across fields from neuroscience to climate science~\cite{Lel_2015,Cohen_2017}, and here we extend their application to urban air quality dynamics.

Using a multi-year dataset covering multiple Indian cities, we analyze interactions between the primary pollutants ($\text{PM}_{2.5}$, $\text{PM}_{10}$, $\text{NO}_{2}$, and $\text{SO}_{2}$) and two of the most important meteorological variables (relative humidity RH and ambient temperature AT). We compute the Pearson correlation coefficient to estimate the linear correlations and use entropy-based measures to quantify uncertainty, dependency, and the directional flow of information~\cite{Pompe_2011,Granger_2009}, thereby capturing  both linear and non linear relationships. 

The resulting correlation data across cities and variable pairs form a high-dimensional dataset that is complex to interpret directly due to the volume of observations and the multiple metrics involved. To systematically extract dominant patterns and reduce dimensional complexity, we amalgamate  these diverse metrics into a unified measure of correlation. This is accomplished by  employing  Principal Component Analysis (PCA)~\cite{jolliffe2016principal,bro2014principal} to the set of correlation measures, enabling a consistent and comparable representation of pollutant–meteorology relationships across cities.  
PCA allows us to construct composite indices that summarize the overall strength and structure of pollutant–weather interactions, both within and across cities, by combining information from all three correlation metrics. 
To the best of our knowledge, this work presents the first implementation of a PCA-based compound indexing methodology for merging multi-metric correlation data between air pollutants and meteorological variables across spatial and variable dimensions.
 
Our results reveal spatial heterogeneity as well as similarities in pollutant–meteorological couplings, suggesting that regional climate and geography play a fundamental role in shaping these dynamics. We find cities cluster into distinct groups with varying levels of interdependence. Additionally, the transfer entropy (TE) analysis provides clear evidence of bidirectional information flow across most cities. In particular, relative humidity (RH) generally exhibits a leading influence on the pollutants, whereas ambient temperature (AT) tends to display the opposite, lagging behavior, with respect to most pollutants. The time delayed  mutual information (TDMI) analysis further reveals that, for the majority of cities, the dependence between particulate matter (PM$_{2.5}$ and PM$_{10}$) and RH peaks at zero lag and decays approximately exponentially with increasing lag. This behavior reflects a short memory and limited temporal persistence in the underlying stochastic dynamics. Only in a subset of cities, similar trend is also observed for the gaseous pollutants (NO$_{2}$ and SO$_{2}$) with RH.

The sequential key steps of our multi-stage analysis are summarized in the following technical roadmap, providing a structured overview that guides the detailed methodology in the subsequent sections:
\begin{itemize}
    \item \textbf{Data preparation and preliminary analysis:} The first stage consists of compilation and pre-processing of multi-city pollutant and meteorological time series, including missing-value treatment. We also investigate the temporal trends and nonlinear pollutant–weather interactions at this stage.
    
    
    
    \item \textbf{Static dependency estimation:} At the next stage, mutual correlation among each pollutant–meteorology pair is estimated using three different static measures, namely, Pearson correlation, mutual information, and relative conditional entropy.
    
    \item \textbf{Intra-city integration:} A Local Correlation Strength for each city is estimated by combining the multiple dependency measures using principal component analysis for all variable pairs.
    
    \item \textbf{Cross-city aggregation (Composite Score $C$):} The spatial variability of the pollutant-weather correlation is investigated at two levels. A Spatial Correlation Score (SCS) is computed using PCA for each variable pair across all the cities. Finally, once again we use PCA to determine a comprehensive correlation score (CCS) combining the SCS across all the variable pairs and all cities. The overall pollutant–meteorology coupling strength $\mathcal{C}$  enables the classification of cities into distinct interaction regimes.

    
    \item \textbf{Temporal dependency analysis:} The mutual temporal dependency of pollutant-weather variable pairs is estimated using transfer entropy (TE) and time-delayed mutual information (TDMI). This analysis is used to characterize causal directionality, lag effects, and persistence in pollutant–weather interactions.
\end{itemize}


The remainder of the paper is organized as follows: Sec.~\ref{sec:data} describes the datasets and methods used in this study, including preliminary analyses of temporal trends. Sec.~\ref{sec:res} presents the results of correlation analyses, PCA-based compound indexing, and temporal dependency assessments. Sec.~\ref{sec:disc} interprets these results, highlighting their implications and limitations. Finally, Sec.~\ref{sec:concl} provides concluding remarks and discusses potential future directions.

\section{Data and Methods} \label{sec:data}

This section describes the datasets and methods used in the study. 

\subsection{Study Location, Sources and Collection Period}
The data used in this study is obtained from the Central Pollution Control Board (CPCB) of India~\cite{1}, which provides publicly available, quality-assured environmental monitoring data. The dataset comprises daily average concentrations of various atmospheric pollutants and meteorological variables collected from various monitoring stations ($N_m$), which recorded the required set of variables in each location across multiple cities [see Table~\ref{tab:1} in Appendix for a summary]. To be precise, data is collected from $87$ monitoring stations, across $24$ cities~\cite{2}, including seven Tier-1 cities (characterized by high population density, robust infrastructure and economic activity) and seventeen Tier-2/3 cities (smaller urban centers with growing population and developing infrastructure).  
The selected cities represent the geographic, climatic, demographic, and industrial diversity of India, encompassing a range of population densities, emission sources, and meteorological conditions~\cite{3,4,5}. The choice is also  based on the availability of continuous, high-quality air pollution  and meteorological data. While not exhaustive, the 87 monitoring stations across these cities provide sufficient coverage to capture representative national-scale air quality dynamics and allow a comprehensive comparative study of pollutant–meteorology interactions.\\

\noindent{\textbf{Selected Environmental Parameters}}: In our analysis, we  focus on a set of key variables representing  both air pollutants and meteorological factors. Specifically, we select $\text{PM}_{2.5}$ (particulate matter with diameter $< 2.5 \,\mu m$), $\text{PM}_{10}$ (particulate matter with diameter $< 10 \, \mu m$), $\text{NO}_{2}$ (nitrogen dioxide) and $\text{SO}_{2}$ (sulfur dioxide) due to their known impact on air quality and human health~\cite{WHO2021AQG}. Among these, PM$_{2.5}$ and PM$_{10}$ are categorized as particulate (matter-type) pollutants, while NO$_2$ and SO$_2$ are gaseous pollutants. In addition two meteorological parameters— RH (relative humidity, expressed as a percentage, representing the amount of moisture in the air relative to the maximum it can hold at a given temperature) and AT (ambient temperature, measured in degrees Celsius, indicating the surrounding air temperature) are included, given their influence~\cite{Hanel_2008,Vaishali_2023} on pollutant dispersion and chemical transformation in the atmosphere. 
Relative humidity (RH) affects aerosol hygroscopic growth, altering particle size, light scattering, and deposition rates, and thereby directly modulating PM$_{2.5}$ concentrations. Ambient temperature (AT) controls  atmospheric stability, influencing both the dilution of pollutants and the rates of secondary aerosol formation. Together, RH and AT capture key thermodynamic and dynamical processes that jointly govern short-term variability in particulate matter.

The choice of these variables is guided by both data availability and their relevance to the study objectives.
\begin{figure*}[tbh]
    \centering
    \includegraphics[width=12cm]{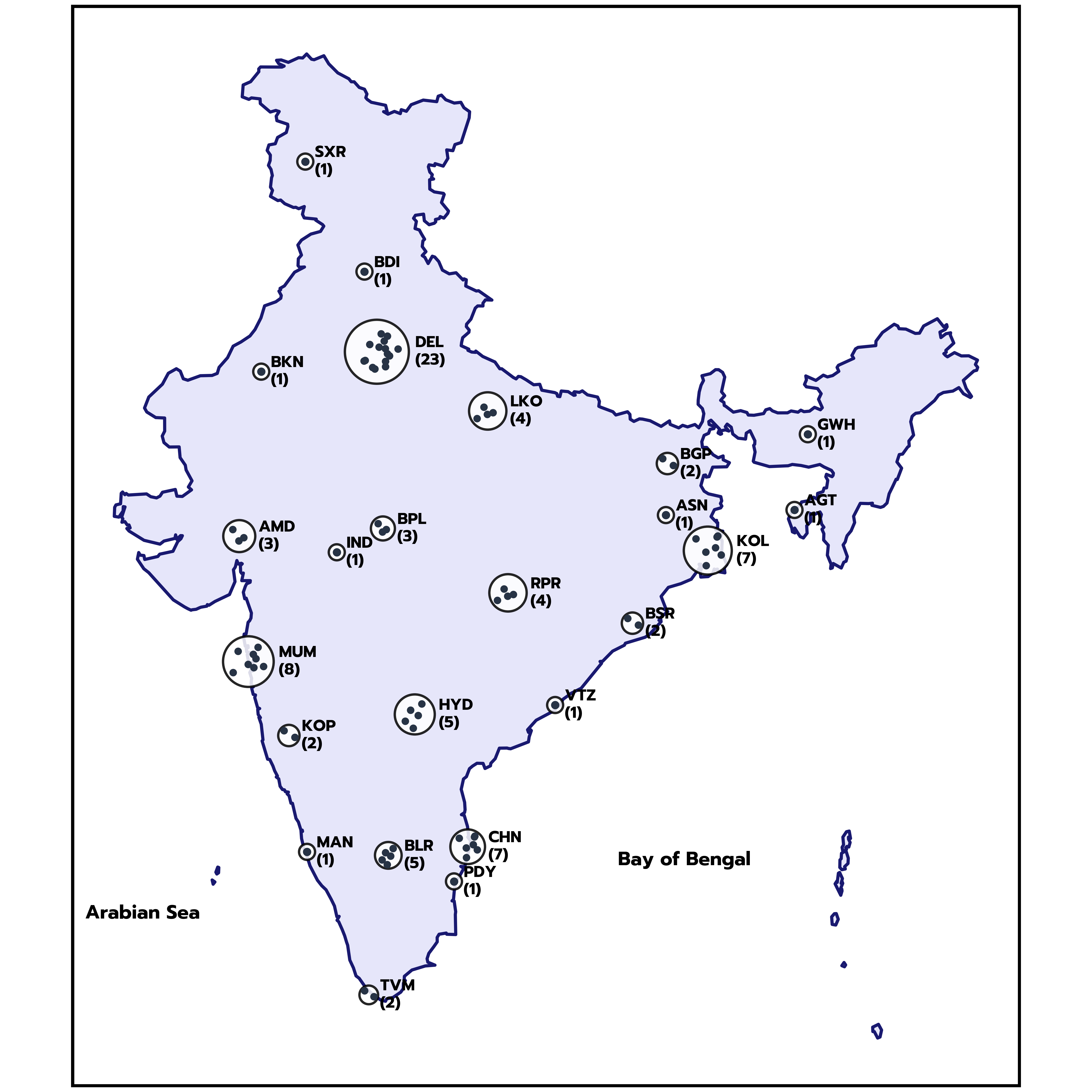}
    \caption{Map of India depicting monitoring stations across the studied cities. Individual stations are shown as dots, and the total number of stations is indicated for each city.}
    \label{fig:map}
\end{figure*}
\subsection{Data Pre-processing }
Environmental data often suffer from missing values and sensor noise due to hardware failures, power outages, or maintenance activities. To address this issue and to smooth the time series for reliable analysis, we employ Kalman filtering~\cite{Kalman,becker2024kalman}, a recursive optimal estimation technique widely used for noisy time series data. For the  datasets used in this study, the average proportion of missing observations is relatively low, amounting to  approximately 2.15\% for PM\textsubscript{2.5}, 3.27\% for PM\textsubscript{10}, 2.41\% for SO\textsubscript{2}, and 2.47\% for NO\textsubscript{2}. Meteorological variables exhibit slightly higher missing rates, with relative humidity (RH) and Ambient Temperature (AT) having approximately 4.08\% and 5.23\% missing values, respectively.
We apply the Kalman filter~\cite{kl} separately to each time series (for all the pollutants and meteorological parameters) for each city. The filter not only interpolates missing values but also de-noises the observed signals, preserving the underlying dynamics critical for entropy and information-theoretic analyses.

\begin{figure*}[tbh]
    \centering    \includegraphics[width=7.5 cm]{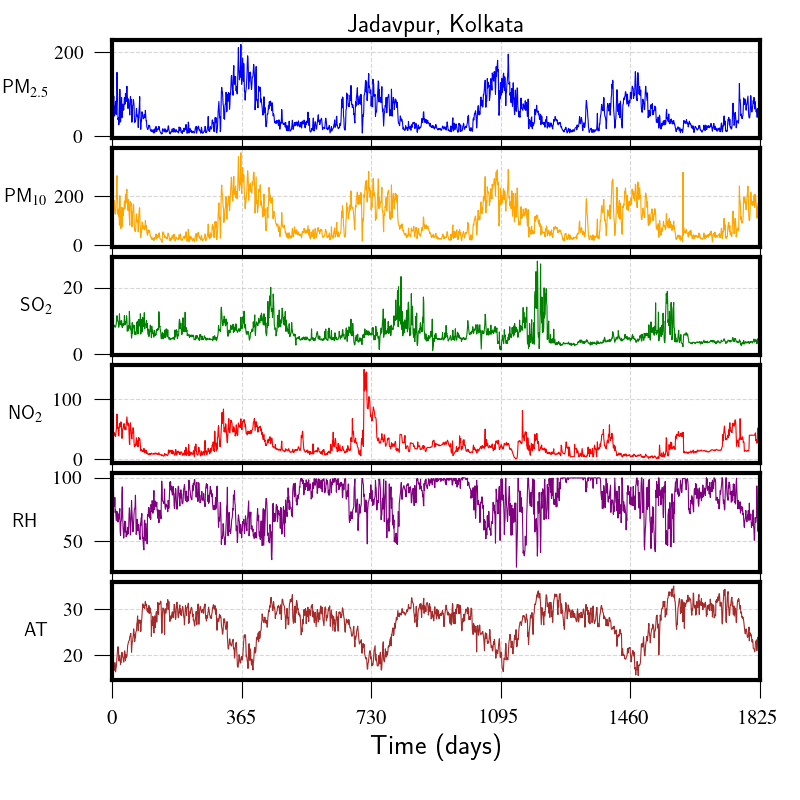}    \includegraphics[width=7.5 cm]{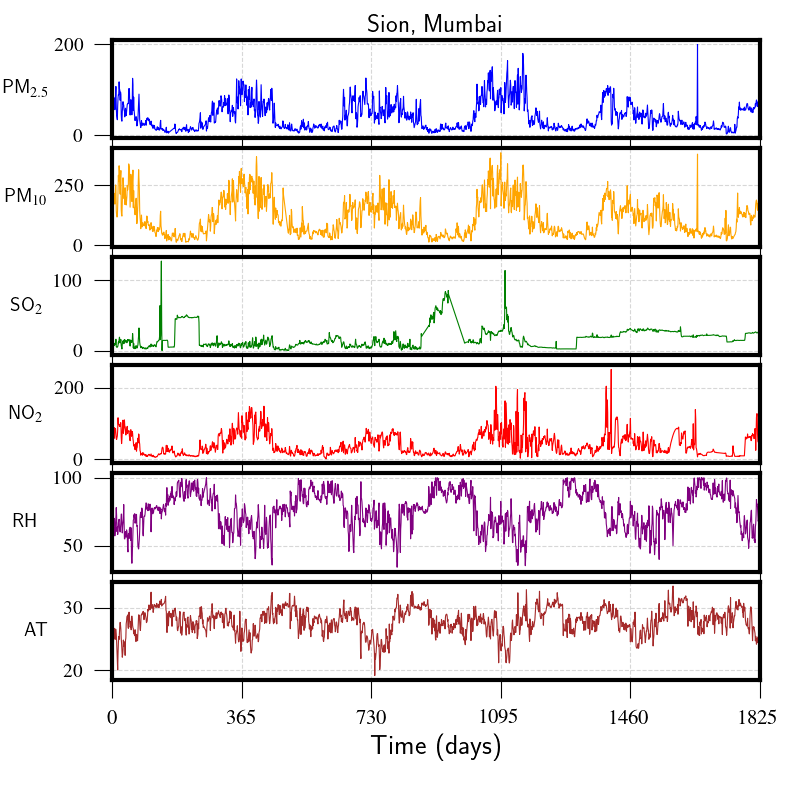}
    \caption{Time-series plots for concentration of the pollutants and the meteorological parameters for two representative stations in Kolkata (left panel) and Mumbai (right panel) for the period Jan, 2020 to Dec, 2024.}
    \label{fig:time_series}
\end{figure*}

The effectiveness of missing value imputation using Kalman Filtering is evaluated across all monitoring stations in 24 cities. To this end, a sensitivity analysis is conducted by artificially removing a subset of observed data and reconstructing them using the Kalman filter for four air pollutants (PM\textsubscript{2.5}, PM\textsubscript{10}, NO\textsubscript{2}, and SO\textsubscript{2}) and two meteorological variables (AT and RH). The imputation accuracy is quantified using the root mean square error (RMSE)~\cite{rmse1} between the observed and reconstructed values.  Consistently low and stable RMSE values across all cities and variables confirm  that  Kalman filtering is an effective and robust method for handling missing observations in our study. 

\subsection{Preliminary Analysis}\label{sec:prelim}
This study employs a multi-layered, entropy-based approach to investigate the interdependence and directional relationships between air pollutants and meteorological variables in Indian cities.
As a starting point, a preliminary analysis is conducted to explore the complex temporal behavior of key pollutants and meteorological parameters. This includes generating time series plots and applying non-parametric regression techniques to identify underlying trends and patterns across different regions. To provide a baseline understanding and quantitative overview of the data, we present the mean ($\langle\rho\rangle$) and standard deviation ($\sigma$) of all pollutants and meteorological variables considered in this study obtained by averaging them across stations in Table~\ref{summary} (see Appendix).

\subsubsection{Temporal Trends}
To visually examine temporal fluctuations and possible seasonal patterns, time series plots of the  pollutant concentrations  and meteorological variables are generated for each location. Fig.~\ref{fig:time_series} depicts such time series plots for two representative stations, namely, Jadavpur in Kolkata and Sion in Mumbai. 

Each panel stacks five years of daily data for six variables - PM\textsubscript{2.5}, PM\textsubscript{10}, SO\textsubscript{2}, NO\textsubscript{2}, RH and AT, so that seasonal patterns amid stochastic fluctuations can be compared at a glance. Both PM\textsubscript{2.5} and PM\textsubscript{10} surge each winter reaching their peaks around Dec-Jan, and decrease during the warm, humid monsoon months pointing to strong seasonality and common sources.
Both the weather variables RH and AT show an almost reverse trend, they reach their minimum during Dec-Jan and become maximum in the summer and monsoon months. 
SO\textsubscript{2} remains near background levels ($<10$ $\mu$g m$^{-3}$) except for a few short spikes, while NO\textsubscript{2} is likewise low save for few dramatic spikes above $100$, reflecting intermittent fluctuations probably driven by local events. 
From these plots for each city, a clear trend emerges: PM\textsubscript{2.5} and PM\textsubscript{10} levels show a strong positive correlation, indicating that they tend to rise and fall together. In contrast, PM\textsubscript{2.5} levels (also the other pollutants) appear to be inversely related to relative humidity (RH), suggesting an anti-correlation where higher humidity is generally associated with lower PM\textsubscript{2.5} concentrations.
\begin{figure*}[tbh]
\centering    
\includegraphics[width=0.92\linewidth]{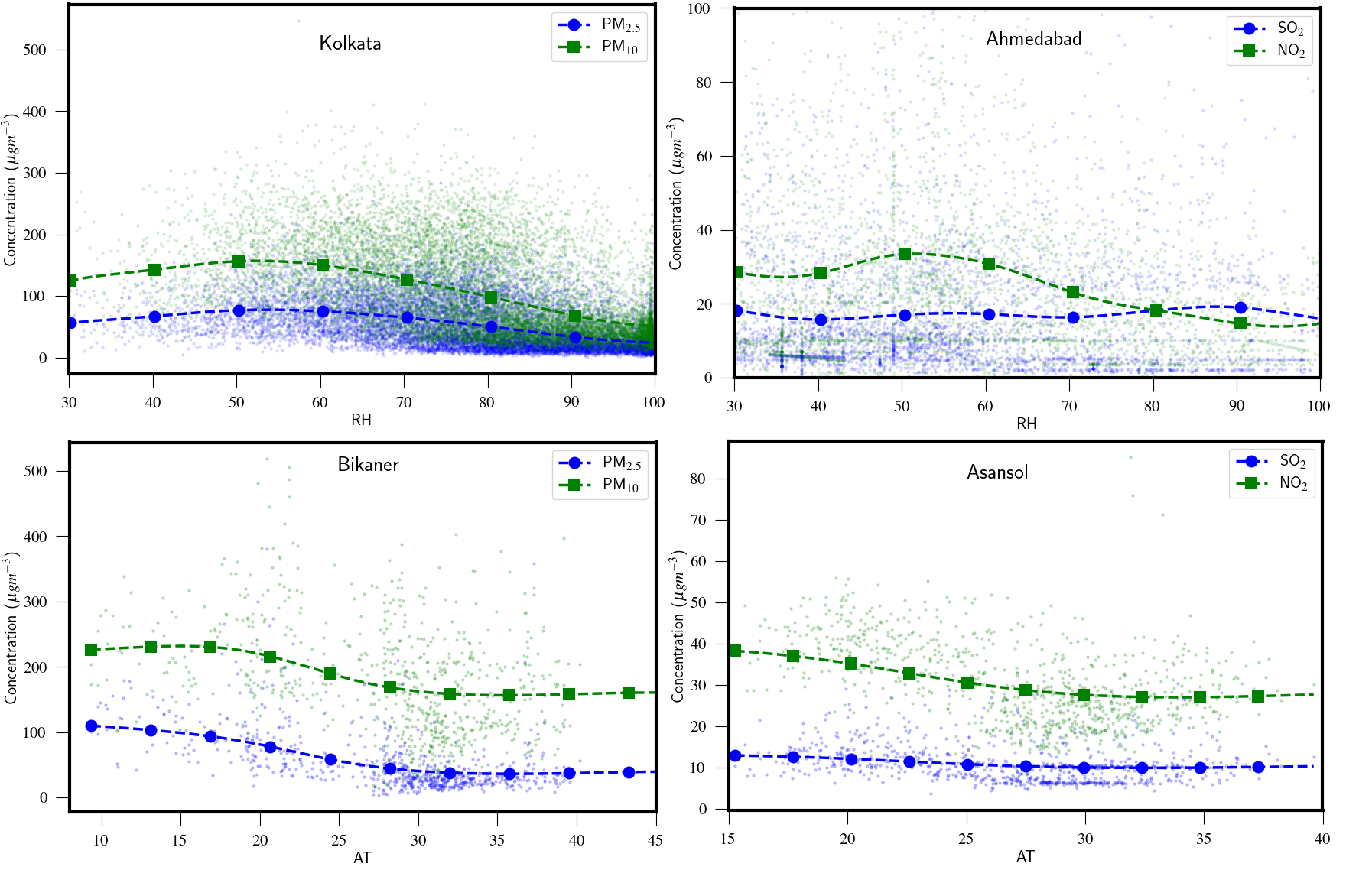} 
\caption{Kernel regression plots illustrating relationships between pollutants and meteorological variables in selected  cities.}
\label{fig:non_params}
\end{figure*}
\subsubsection{Nonlinearity in Correlation}
To complement the time-series diagnostics, we next explore the instantaneous relationships between each air-quality indicator (PM\textsubscript{2.5}, PM\textsubscript{10}, NO\textsubscript{2}, SO\textsubscript{2}) and the two meteorological drivers (RH, AT). For every pollutant–weather variable pair we pool the full five-year daily dataset and fit a non-parametric regression smoothing with a kernel regression framework~\cite{non_param1}.
In statistics, kernel regression is a non-parametric method used to detect the type of relationship present between random variables $X$ and $Y$. The regression takes the form,
\begin{equation}
    E(Y|X)=m(X)
\end{equation}
where $E(Y|X)$ denotes the expectation value of $Y$, conditioned on $X$, and $m$ is an unknown function. A common estimator for this function is the Nadaraya–Watson estimator,
\begin{equation}
    \hat{m}(x)=\frac{\sum_{i=1}^NK_h(x-X_i)Y_i}{\sum_{i=1}^NK_h(x-X_i)},
\end{equation}
with $Y_i$ as the pollutant concentration, $X_i$ the weather variable, and $K_h(z)$ is a Gaussian kernel with bandwidth $h$ selected via leave-one-out cross-validation~\cite{non_param,non_param2}. This approach places more weight on observations whose meteorological conditions are close to the point of interest, without imposing a parametric shape, thereby naturally handling fluctuations across scales.

Figure~\ref{fig:non_params} show the kernel regression plots for different pairs of pollutant-meteorological parameters for a few representative cities. 

These plots demonstrate clear nonlinear relationships between the pollutants and meteorological variables, indicating that traditional linear methods may not adequately assess  the complexity of these interactions.


\subsection{Correlation Analysis of Pollutants and Weather variables }\label{sec:corr}

In light of the nonlinear relationships identified in the preliminary analysis, the subsequent analysis focuses on systematically characterizing the associations among air pollutants and meteorological parameters.
To ensure a robust and comprehensive characterization of these relationships, we employ a multi metric approach that accounts for diverse interaction patterns. We focus on  three complementary correlation metrics: Pearson coefficient, mutual information, and relative conditional entropy. While the Pearson coefficient estimates the linear dependencies, the other two entropy based measures are more general, and are able to characterize non-linear and asymmetric dependencies, as well as variability arising from fluctuations across timescales. 

We now discuss the methods used to compute each of the three metrics in details to understand their  individual pollutant-weather relationships  and lay the groundwork for the construction of a unified metric  in a later section.

\subsubsection{Pearson Correlation Coefficient}
The simplest quantitative estimate of the linear interdependency of two observables is the Pearson correlation coefficient (PCC)~\cite{gun2008fundamentals}. 
The PCC between two time series $X=\{x_{1},x_{2},..., x_{N}\}$ and $Y=\{y_{1},y_{2},...,y_{N}\}$ is defined as,
\begin{align} \label{eqn:PCC}
  r_{X,Y}=\frac{\sum_{t=1}^{N}(x_t-\hat{x})(y_t-\hat{y})}{\sqrt{\sum_{t=1}^{N}(x_t-\hat{x})^{2}}\,\sqrt{\sum_{t=1}^{N}(y_t-\hat{y})^{2}}},  
\end{align}
where, $\hat{x}=\frac {1}{N} \sum_{t=1}^{N}x_t$ and $\hat{y}=\frac {1}{N} \sum_{t=1}^{N}y_t$ are the means of the respective time-series. 
Here $N$ denotes the length of the time-series, which, for our case varies from station to station. Clearly, PCC is symmetric under the exchange of $X$ and  $Y$ and its values are bounded by $-1 \le r_{X,Y} \le 1$. While a perfect positive (negative) linear correlation between $X$ and $Y$ is indicated by $r_{X,Y}=1$ ($r_{X,Y}=-1$), $r_{X,Y}=0$ indicates absence of linear correlation.  

For each of the selected cities, we compute the PCC between the eight pollutant--weather variable pairs, namely, PM\textsubscript{2.5}-RH, PM\textsubscript{2.5}-AT, PM\textsubscript{10}-RH etc. To this end, we first compute the PCC value $r_{X,Y}^k$ between the pollutant $X$ and weather variable $Y$ from the corresponding time-series data from the $k$-th station. The PCC $r_{X,Y}$ for the entire city is then obtained by averaging over all the stations in the city,
\begin{equation} \label{eqn:4}
    r_{X,Y}=\frac  1 {N_m} \sum_{k=1}^{N_m}r_{X,Y}^k,
\end{equation}
where $N_m$ denotes the total number of monitoring stations in a city. 

\subsubsection{Mutual Information}
To further explore the dependence structure between air pollutants and weather variables, we compute the mutual information~\cite{Steu_2002,Kras_2004}. 
MI quantifies the reduction in uncertainty of one variable given knowledge of the other, characterizing both linear and non-linear dependencies. The mutual information  between air pollutants ($X$) and meteorological factors ($Y$) is defined as 
 \begin{equation} \label{eqn:5}
I_{X,Y}=S_X+S_Y-S_{X,Y}\,\,,
\end{equation}
where $S_X$ and  $S_Y$ represent the self-entropy (Shannon Entropy) of variables $X$ and $Y$, respectively, and $S_{X,Y}$ denotes their joint entropy.

Let $p(x)$ denote the probability density function (PDF) of the stationary time-series $X = \{x_t; t=1, \cdots N\}$. The corresponding Shannon entropy, which quantifies the average uncertainty or information content in a PDF, is defined as, 
\begin{equation}
    S_X = -\int dx\,p(x)\,\ln[p(x)]. \label{eq:Sx}
\end{equation} 
Higher entropy values indicate greater  uncertainty  or disorder in the  system. The joint entropy  $S_{X,Y}$, which quantifies the combined uncertainty of both variables, is defined as,
\begin{equation} 
   S_{X,Y}=-\int dx \int dy\,p(x,y)\,\ln[p(x,y)], \label{eq:Sxy}
\end{equation} 
where $p(x,y)$ denotes the joint probability density of $X$ and $Y$.
 
We construct the  probability distribution function (PDF) of each observable (all pollutants and weather variables) from the aggregated data from all the stations in each city. The joint distributions of each pollutant-weather variable pair is also constructed from the same data. The self-entropies of each observable and the mutual information of the pollutant-weather variables pairs are computed from these distributions. 

\subsubsection{Relative Conditional Entropy}
To further estimate the strength, complexity and conditional structure of interactions between air pollutants ($X$) and meteorological factors ($Y$), (both linear and non-linear), we compute relative conditional entropy.
The conditional entropy for a pollutant ($X$), given a weather variable ($Y$) is defined as,
\begin{equation} \label{eqn:8}
   {\cal H}_{X;Y} = S_{X,Y} - S_{Y},
\end{equation}
where $S_{X,Y}$ and $S_Y$ denote the joint entropy of $X$ and $Y$ [see Eq.~\eqref{eq:Sxy}] and self-entropy of $Y$ [see Eq.~\eqref{eq:Sx}], respectively.
We compute the self entropy of each pollutant ($X$) and weather variable ($Y$) across all cities and reported these values in Table~\ref{tab:Entropy} in Appendix.
To characterize the fraction of uncertainty remaining in $X$ after conditioning on $Y$, it is useful to define a relative conditional entropy as,
\bea
{\cal H}^R_{X;Y} = {\cal H}_{X;Y}/ S_X. \label{eqn:9}
\eea 
It can be easily shown that $ 0 \le {\cal H}^R_{X;Y} \le 1$. Smaller values of ${\cal H}^R_{X;Y}$ indicate that $X$ has a strong dependence on $Y$, while 
${\cal H}^R_{X;Y}=1$ indicates complete independence of the two variables. 

\subsection{PCA-based Compound Scoring Framework for Pollutant–Meteorological Correlation Strength} \label{sec:multi} 

The different metrics discussed so far, namely, Pearson correlation, relative conditional entropy and MI,  complement each other in characterizing the interdependence of the pollutants and weather variables.  While individual correlation measures can highlight pairwise relationships between pollutants and meteorological variables, a more comprehensive understanding requires integrating these metrics and scaling the analysis across different spatial levels.  
To develop a unified measure of the correlations, we employ  Principal Component Analysis (PCA)~\cite{jolliffe2016principal,bro2014principal}, a well established technique frequently used in the development of composite indices. PCA provides a rigorous statistical framework for identifying latent structures in multivariate data by projecting it onto orthogonal axes (principal components) that maximize variance. In the context of correlation analysis, PCA allows us to construct composite indices that summarize the overall strength and structure of pollutant–weather interactions, both within and across cities, by combining information from all three correlation metrics. This dimensionality reduction  in multivariate datasets facilitates the development of interpretable and scalable indices that can be used to compare locations, identify regional trends, and support further modeling.

To describe PCA from a general perspective, let us consider a dataset with $p$ features, each measured over 
$m$ observations. For our case, a feature can be a specific correlation measure or a pollutant-weather variable pair, while an observation can be a particular pollutant-weather pair or a city, depending on the context. The data can be represented as a matrix of dimension $m\times p$ where each row corresponds to an observation and each column to a feature. The element $x_{jk}$  denotes the value of feature $k$ for observation $j$. Prior to applying PCA, the data are standardized to ensure all features are on a comparable scale. Each element is transformed to its Z-score form,
\begin{align} \label{eqn:10}
z_{jk} &= \frac{x_{jk} - \mu_k}{\sigma_k}, \nonumber \\
&\quad \forall j = 1, 2, \ldots, m ~ 
\text{and}~  k = 1, 2, \ldots, p,
\end{align}
where $\mu_{k} = \frac 1m \sum_{j=1}^m x_{jk}$ denotes the mean of $k$-th feature over the $m$ observations and $\sigma_k$ denotes the corresponding standard deviation.

This step ensures that all metrics are placed on a common scale, allowing for meaningful combination despite differences in their original units and distributional characteristics. PCA is then performed on the resulting standardized matrix, yielding a set of principal components ordered by the amount of variance they contain. In each  case, we retain only the first principal component (PC1), which contains  the largest share of total variance, and is given by,
\begin{equation} 
\text{PC1}_j = \sum_{k=1}^{p} w_k \cdot z_{jk}, \label{eq:PC1}
\end{equation}
where $w_k$ is the loading (i.e., the weight) of the $k$-th feature. 
The first principal component accounts for the maximum amount of variability of the features compared to all the other principal components formed after the dimensional reduction. Thus, $\text{PC1}_j$ can be interpreted as 
a unified measure for all the features for the $j$-th observation. 

To systematically interpret the complex interdependencies between air pollutants and meteorological variables across multiple locations, through a unified measure of correlation strength, we employ a three-tiered PCA-based correlation aggregation framework.

\begin{itemize}
    \item At the first level, intra-city correlation strengths are computed for each pollutant–weather variable pair, capturing localized interactions within individual cities. In this case, `observation' refers to the eight different pollutant-weather variable pairs, i.e., $m=8$ while `feature' refers to the different measures of correlation, i.e., $p=3$.

\item The second level involves inter-city aggregation, wherein correlation strengths for each variable pair are consolidated across cities to reveal broader spatial patterns. In this case, each `observation' corresponds to a city ($m=24$)  whereas each `feature' represents a distinct correlation metric ($p=3$).

\item Finally, these pairwise insights are combined into a comprehensive correlation score (CCS) that encapsulates the overall degree of correlation between pollution and weather correlations across the network. Here the  `observation' refers to the 24 cities ($m=24$) 
and the `features' correspond  to the different pollutant-weather variable pairs ($p=8$).

\end{itemize}
\textbf{Sensitivity Analysis:}To investigate the influence of different pairs of pollutant-weather correlations to the values of CCS across multiple cities, we conduct a PCA-based sensitivity analysis~\cite{jolliffe2011principal,Tanaka01011997,Tanaka1988SensitivityAI}. This analysis quantifies how the exclusion of each individual pollutant–weather pair affects the overall comprehensive correlation score, allowing us to assess the relative contribution of each interaction to the city-level correlation structure.





\subsection{Temporal Correlation}\label{sec:temp}

It is important to go beyond static correlation measures and investigate the temporal dynamics between pollutants and meteorological variables to uncover potential causal linkages and directional dependencies. In this subsection the variable pairs are  subjected to time-domain analysis using two complementary techniques, namely Transfer Entropy (TE) and time-delayed Mutual Information (TDMI). This approach allows us to explore both the directionality and temporal structure of interactions, providing insights into the lead-lag behavior and potential causal links underlying pollutant–weather dynamics.

\subsubsection{Causal Interaction: Transfer Entropy} 

We utilize transfer entropy~\cite{TE_2000,TE_2011}, a non-parametric  information-theoretic measure to explore the directional dependencies and potential causal interactions between air pollutants and meteorological parameters. TE  quantifies the amount of information that past values of a meteorological  variable convey about the future evolution of a pollutant (and vice versa) beyond what is already contained in each series’ own history and thereby indicates possible causal influence.
TE from a process $Y$ to another process $X$ is defined as,
\begin{align} 
{\cal T}_{Y\xrightarrow{}X}&={\cal H}_{X_{t};X_{t-1}}-{\cal H}_{X_{t};X_{t-1},Y_{t-1}}, \label{eq:Txy}
\end{align}
where ${\cal H}_{X_t; X_{t-1}}$ is the conditional entropy of $X_t$, given its recent past value $X_{t-1}$ [see Eq.~\eqref{eqn:8}]. Moreover, ${\cal H}_{X_{t};X_{t-1},Y_{t-1}}$ denotes entropy of $X_t$, conditioned on the past values of both $X$ and $Y$, given by,
\begin{equation}
    {\cal H}_{X_{t};X_{t-1},Y_{t-1}} = S_{X_{t},X_{t-1},Y_{t-1}}-S_{X_{t-1},Y_{t-1}}, \label{eq:Hxyy}
\end{equation}
where $S_{X_{t-1},Y_{t-1}} = S_{X,Y}$ is the joint entropy of $X$ and $Y$, defined in Eq.~\eqref{eq:Sxy}, and
\begin{align}
&  S_{X_{t},X_{t-1},Y_{t-1}}  = -\int dx_t\int dx_{t-1}  \int dy_{t-1}\,\cr 
  & \quad \quad \times \, p(x_t,x_{t-1},y_{t-1})\,\ln p(x_t,x_{t-1},y_{t-1}), \label{eq:Sxyz}
\end{align}
denotes the joint entropy of $X$ with past values of itself and $Y$. Using Eqs.~\eqref{eq:Hxyy}-\eqref{eq:Sxyz}, Eq.~\eqref{eq:Txy} simplifies to,
\begin{align}
{\cal T}_{Y\to X}=S_{X_{t},X_{t-1}}-
 S_{X_{t},X_{t-1},Y_{t-1}}+S_{X,Y}-S_{X}, \label{eq:Tyx_2}
\end{align}
Note that, ${\cal T}_{Y\to X}$ is not symmetric in $X$ and $Y$. A positive ${\cal T}_{Y\to X}$ implies that past values of $Y$ has a directional effect on present values of $X$. This helps in identifying causal relationships, such as whether changes in $Y$ precede changes in $X$ or vice versa. 

\subsubsection{Time-Delayed Mutual Information}
To further explore the temporal dependence between  pollutant and weather variables, we compute the mutual information with a time delay, $\tau$, considering  pollutant at day $t$ and meteorological parameter at a later day $t +\tau$.
For two stationary time-series $X=\{x_t\}_{t=1}^N$ (pollutant) and $Y=\{y_t\}_{t=1}^N$ (weather variable), TDMI for a given lag $\tau$ is defined as,
\begin{align} \label{eqn:23}
I_{X,Y}(\tau) = \sum_{x_t}\sum_{y_{t+\tau}} p(x_t,y_{t+\tau})\,
\ln\left[\frac{p(x_t,y_{t+\tau})}{p(x)p(y)}\right],
\end{align}
where $p(x_t,y_{t+\tau})$ is the joint probability distribution of $(X_t,Y_{t+\tau})$ and $p(x)$, $p(y)$ are the corresponding marginals. Equivalently, in terms of entropies, we can write,
\begin{align} \label{eqn:24}
I_{X,Y}(\tau) = S_{X} + S_{Y} - S_{X_t,Y_{t+\tau}},
\end{align}
with $S_{X_t,Y_{(t+\tau)}}$ denoting the joint entropy.
TDMI thus quantifies lagged statistical dependencies between pollutant and weather variables.

\section{Results} \label{sec:res}

In this section, we present the main findings of our study, focusing on static and dynamic interdependencies of air pollutants and meteorological variables across the study regions. The results are organized into three main subsections. First, we analyze the static inter-dependencies among the variables. Next, we compute the correlation score depending on PCA based framework for intra-city as well as inter-city aggregation. Finally, we explore the temporal dynamics of the variables to understand the directional dependencies between variable pairs.

\subsection{Static Dependency Analysis}
This subsection examines the static relationships between air pollutants and meteorological variables. The main findings for three complementary measures employed to quantify these associations are discussed below.

\subsubsection{Pearson Correlation}
\begin{table*}[th]
\centering
\small
\begin{tabular}{|c||c|c|c|c|l|l|c|c|}
\hline
\textbf{City} & $r_{\text{\tiny PM}_{2.5},\text{\tiny RH}}$ & $r_{\text{\tiny PM}_{2.5},\text{\tiny AT}}$ & $r_{\text{\tiny PM}_{10},\text{\tiny RH}}$ & $r_{\text{\tiny PM}_{10},\text{\tiny AT}}$  & $r_{\text{\tiny SO}_{2},\text{\tiny RH}}$ 
&$r_{\text{\tiny SO}_{2},\text{\tiny AT}}$ 
& $r_{\text{\tiny NO}_{2},\text{\tiny RH}}$ & $r_{\text{\tiny NO}_{2},\text{\tiny AT}}$ \\\hline
AMD & -0.278 & -0.163 & -0.305 & -0.151  & -0.151 
&-0.020 
& -0.012 & -0.027 \\\hline
BLR & -0.191 & \cellcolor{blue!25}{0.038} & -0.312 & \cellcolor{blue!25}{0.116}  & -0.188 
&-0.047 
& \cellcolor{blue!25}{0.044} & \cellcolor{blue!25}{0.016} \\\hline
CHN & \cellcolor{blue!25}{0.044} & -0.290 & -0.084 & -0.222  & -0.025 
&-0.139 
& \cellcolor{blue!25}{0.051} & -0.182 \\\hline
DEL & \cellcolor{blue!25}{0.189} & -0.562 & -0.085 & -0.407  & -0.062 
&-0.338 
& -0.331 & -0.005 \\\hline
HYD & -0.394 & -0.171 & -0.497 & -0.060  & -0.283 
&-0.093 
& \cellcolor{blue!25}{0.137} & \cellcolor{blue!25}{0.173} \\\hline
KOL & -0.425 & -0.522 & -0.476 & -0.503  & -0.283 
&-0.283 
& -0.415 & -0.172 \\\hline
MUM & -0.544 & -0.214 & -0.561 & -0.170  & -0.473 
&-0.182 
& \cellcolor{blue!25}{0.023} & -0.001 \\\hline
AGT & \cellcolor{blue!25}{0.098} & -0.239 & -0.019 & -0.284  & \cellcolor{blue!25}{0.033} 
&-0.154 
& -0.158 & -0.297 \\\hline
ASN & -0.353 & -0.491 & -0.385 & -0.534  & -0.431 
&-0.486 
& -0.507 & -0.280 \\\hline
BDI & \cellcolor{blue!25}{0.088} & -0.578 & -0.128 & -0.373  & \cellcolor{blue!25}{0.102} 
&-0.621 
& -0.036 & -0.410 \\\hline
BGP & -0.032 & -0.633 & -0.238 & -0.381  & -0.238 
&-0.275 
& \cellcolor{blue!25}{0.137} & \cellcolor{blue!25}{0.047} \\\hline
BPL & -0.188 & -0.536 & -0.393 & -0.373  & -0.509 
&-0.350 
& -0.226 & -0.009 \\\hline
BSR & -0.295 & -0.661 & -0.340 & -0.568  & -0.214 
&-0.758 
& -0.302 & -0.145 \\\hline
BKN & \cellcolor{blue!25}{0.009} & -0.544 & -0.408 & -0.357  & -0.317 
&-0.348 
& -0.270 & -0.338 \\\hline
GWH & -0.077 & -0.482 & -0.122 & -0.422  & \cellcolor{blue!25}{0.018} 
&-0.300 
& \cellcolor{blue!25}{0.312} & \cellcolor{blue!25}{0.117} \\\hline
IND & -0.197 & -0.304 & -0.505 & -0.106  & -0.537 
&-0.176 
& -0.485 & \cellcolor{blue!25}{0.128} \\\hline
KOP & -0.487 & \cellcolor{blue!25}{0.017} & -0.618 & \cellcolor{blue!25}{0.067}  & -0.609 
&\cellcolor{blue!25}{0.054} 
& -0.416 & \cellcolor{blue!25}{0.070} \\\hline
LKO & -0.093 & -0.376 & -0.331 & -0.272  & -0.188 
&-0.165 
& -0.101 & \cellcolor{blue!25}{0.035} \\\hline
MAN & \cellcolor{blue!25}{0.085} & \cellcolor{blue!25}{0.001} & \cellcolor{blue!25}{0.135} & -0.013  & -0.468 
&\cellcolor{blue!25}{0.303} 
& -0.366 & -0.090 \\\hline
PDY & -0.023 & -0.303 & -0.297 & -0.095  & -0.011 
&-0.439 
& -0.247 & \cellcolor{blue!25}{0.248} \\\hline
RPR & -0.381 & -0.294 & -0.435 & -0.254  & -0.545 
&-0.125 
& -0.422 & \cellcolor{blue!25}{0.293} \\\hline
SXR & -0.100 & \cellcolor{blue!25}{0.007} & -0.018 & -0.067  & -0.028 
&-0.246 
& -0.086 & -0.096 \\\hline
TVM & -0.327 & \cellcolor{blue!25}{0.120} & -0.366 & \cellcolor{blue!25}{0.211}  & \cellcolor{blue!25}{0.006} 
&\cellcolor{blue!25}{0.079} 
& \cellcolor{blue!25}{0.009} & \cellcolor{blue!25}{0.128} \\\hline
VTZ & -0.175 & -0.434 & -0.157 & -0.205  & -0.151 &-0.101 & \cellcolor{blue!25}{0.158} & -0.032 \\\hline
\end{tabular}
\vspace{0.2 cm}
\caption{Pearson Correlation Coefficients of pollutants and weather variables across cities with maximum $7\%$ margin of error. Blue colored cells highlight positive correlation.}
\label{tab:pearson}
\end{table*}
The PCC values computed from Eqs.~\eqref{eqn:PCC} and \eqref{eqn:4} for the twenty four selected cities are  summarized in Table \ref{tab:pearson}. 
It appears that $r_{\text{PM}_{2.5},\text{RH}}<0$ for most of the cities, which indicates that PM\textsubscript{2.5} and relative humidity affect each other negatively, at least to linear order. A similar trend is observed for PM\textsubscript{10}, with RH showing a negative correlation with it in most locations, reinforcing the idea that moisture in the air can suppress particulate matter levels. Ambient temperature (AT), too, tends to correlate negatively with PM\textsubscript{2.5} and PM\textsubscript{10} across many cities, especially in urban centers like Kolkata, Delhi, Mumbai and Asansol, where cooler temperatures may coincide with poor dispersion conditions or increased local emissions, leading to pollutant accumulation. For gaseous pollutants like SO\textsubscript{2} and NO\textsubscript{2}, the relationships are more varied. While several cities show weak or inconsistent correlations, others (e.g., Kolkata, Asansol, Bhubaneswar, and Bikaner) display significant negative correlations. Overall, both RH and AT have a measurable influence on air pollutant concentrations, particularly for particulate matter, and that these meteorological variables should be considered when analyzing or forecasting urban air quality.

\subsubsection{Mutual Information}
\begin{table*}[th]
    \centering
   { \small
    \begin{tabular}{|c||c|c|c|c|c|c|c|c|}
\hline
{\textbf{City}} & $I_{\text{\tiny PM}_{2.5},\text{\tiny RH}}$ & $I_{\text{\tiny PM}_{2.5},\text{\tiny AT}}$ & $I_{\text{\tiny PM}_{10},\text{\tiny RH}}$ & $I_{\text{\tiny PM}_{10},\text{\tiny AT}}$ & $I_{\text{\tiny SO}_{2},\text{\tiny RH}}$ & $I_{\text{\tiny SO}_{2},\text{\tiny AT}}$ & $I_{\text{\tiny NO}_{2},\text{\tiny RH}}$ & $I_{\text{\tiny NO}_{2},\text{\tiny AT}}$ \\
\hline
AMD&0.180 &0.144 &0.196 &0.151 &0.153 & 0.106& 0.162& 0.136 \\\hline
 BLR& \cellcolor{red!25}0.073& \cellcolor{red!25}0.023& 0.117& \cellcolor{red!25}0.059& 0.064& 0.030& 0.097&\cellcolor{red!25}0.047
\\
    \hline
    CHN&  0.076& 0.110& \cellcolor{red!25}0.054&  0.112&  0.082& 0.088& 0.069& 0.091
\\
    \hline
    DEL& 0.105& 0.240& 0.092& 0.139& 0.060& 0.022& \cellcolor{red!25}0.017& 0.078
\\
    \hline
    HYD& 0.204& 0.073& 0.234& 0.076& 0.082& 0.127& 0.128& 0.111
\\
    \hline
    KOL& 0.212& 0.265& 0.221& 0.269& 0.043& \cellcolor{red!25}0.018& 0.082& 0.104
\\
    \hline
    MUM& 0.268& 0.071& 0.305& 0.061& \cellcolor{red!25}0.031& 0.080& 0.166& 0.067
\\
    \hline
    AGT& 0.400& 0.515& 0.360& 0.554& 0.272& 0.406& 0.389& 0.502
\\
   \hline
    ASN& \cellcolor{blue!25}0.597&0.602 & 0.599&0.570 &\cellcolor{blue!25}0.555 &0.534&0.475&0.536 \\
    \hline
    BDI&0.329 &0.516 &0.347 & 0.439&0.456 &0.499 &0.357 &0.531 \\
    \hline
    BGP&0.280 &0.414 & 0.279&0.296 & 0.120& 0.156&0.215 & 0.289\\
    \hline
    BPL& 0.333& 0.380& 0.381& 0.332&0.223& 0.256&0.403&0.361
\\\hline
BSR& 0.496&\cellcolor{blue!25}0.726&0.488&\cellcolor{blue!25}0.653&0.274&0.355&0.504&\cellcolor{blue!25}0.755
\\
        \hline
    BKN&0.521 & 0.559&\cellcolor{blue!25}0.703 &0.535&0.333&0.350&0.530&0.499\\
        \hline
    GWH& 0.348& 0.470& 0.385& 0.469& 0.474&0.439&0.242&0.325
\\
        \hline
    IND&0.244&0.241&0.451&0.304&0.254&0.159&0.406&0.308
\\
        \hline
    KOP&0.471 &0.196 &0.617&0.310&0.240&0.086&\cellcolor{blue!25}0.553&0.246\\
        \hline
    LKO&0.145 &0.198&0.180&0.160&0.203&0.184&0.118&0.103\\
        \hline
    MAN&0.311 & 0.230& 0.339& 0.293&0.512& \cellcolor{blue!25}0.599&0.297&0.309\\
        \hline
    PDY&0.208 &0.229 & 0.301&0.245 &0.260&0.208 &0.208&0.306\\
        \hline
    RPR& 0.253&0.163&0.301&0.220&0.184&0.195&0.317&0.206
\\
        \hline
    SXR&0.222&0.272&0.247&0.298&0.176&0.293&0.312&0.450\\
\hline
    TVM&0.183& 0.253&0.144&0.253&0.263&0.416&0.110&0.158
\\
        \hline
 VTZ
& 0.196& 0.292& 0.209& 0.267& 0.162& 0.166& 0.162&0.216
\\ \hline 
    \end{tabular}
   }
   \vspace{0.2 cm}
    \caption{Mutual Information values for the eight pollutant-meteorological variable pairs across the selected cities. For each pair, the maximum and minimum MI values are highlighted in blue and red, respectively.}
    \label{tab:MI}
\end{table*}
Mutual Information ($I_{X,Y}$) values are computed using the self entropy [see Eq.~\eqref{eq:Sx}] and joint entropy [see Eq.~\eqref{eq:Sxy}] of the corresponding pollutant and meteorological variables. 
Table~\ref{tab:Entropy} (Appendix) summarizes the self-entropy values indicating their individual uncertainty levels.
It is apparent from this table that across cities the uncertainty ranking is roughly $S_{\text{PM}_{10}}\geq S_{\text{PM}_{2.5}}>S_{\text{NO}_{2}}\sim S_{\text{RH}}>S_{\text{SO}_{2}}>S_{\text{AT}}$, highlighting that particulates fluctuate most from day to day, while temperature is comparatively stable. 

The total amount of  shared information between pollutant and meteorological variables is estimated by  computing  $I_{X,Y}$  using Eq.~\eqref{eqn:5}, \eqref{eq:Sx} and \eqref{eq:Sxy} for all variable pairs, across all  cities.  
Table \ref{tab:MI} summarizes MI between each pollutant and the two meteorological drivers (RH, AT) for all cities.
Broadly, particulates show the strongest weather links, often tighter with humidity than temperature: e.g. $I$\textsubscript{PM,RH} exceeds $I$\textsubscript{PM,AT} in Ahmedabad, Bengaluru, Hyderabad, Mumbai, Kolhapur, etc., while a few places (Chennai, Kolkata) show comparable or higher $I$\textsubscript{PM,AT}. Several cities stand out with consistently high $I$ across many pairs notably Asansol (lying in the range $0.47-0.60$ for most pairs), Agartala, Baddi, Bhubaneswar, Bikaner, and Kolhapur for $I$\textsubscript{PM\textsubscript{10},RH} ($0.617$). The largest value of $I$ in the table is $I$\textsubscript{NO\textsubscript{2},AT} in Bhubaneswar ($0.755$), indicating strong temperature control on NO\textsubscript{2} there; $I$\textsubscript{PM\textsubscript{10},RH} in Bikaner ($0.703$) and PM pairs in Asansol ($0.60$) are also prominent. For SO\textsubscript{2}, the strongest dependencies tend to be with AT in certain climates e.g., Mangaluru ($0.599$), Thiruvananthapuram ($0.416$), Srinagar ($0.293$) suggesting thermally driven chemistry or sources, whereas other cities show modest SO\textsubscript{2}–weather coupling. Coastal metros like Mumbai display higher $I$\textsubscript{PM,RH} ($0.268-0.305$) but weaker $I$\textsubscript{PM,AT}, consistent with moist boundary-layer effects. Overall, $I$ values are city-specific, but the common pattern is $I_{\text{PM} \leftrightarrow \text{RH}} > I_{\text{PM} \leftrightarrow \text{AT}}$, while NO\textsubscript{2} often tracks AT, and SO\textsubscript{2} shows mixed but sometimes temperature-led behavior.

\subsubsection{Relative Conditional Entropy}
\begin{table*}[th]
\centering
 \small
    \begin{tabular}{|c||c|c|c|c|c|c|c|c|}
\hline
{\textbf{City}} & ${\cal H}^R_{\text{\tiny PM}_{2.5};\text{\tiny RH}}$ & ${\cal H}^R_{\text{\tiny PM}_{2.5};\text{\tiny AT}}$ & ${\cal H}^R_{\text{\tiny PM}_{10};\text{\tiny RH}}$ & ${\cal H}^R_{\text{\tiny PM}_{10};\text{\tiny AT}}$ & ${\cal H}^R_{\text{\tiny SO}_{2};\text{\tiny RH}}$ & ${\cal H}^R_{\text{\tiny SO}_{2};\text{\tiny AT}}$ & ${\cal H}^R_{\text{\tiny NO}_{2};\text{\tiny RH}}$ & ${\cal H}^R_{\text{\tiny NO}_{2};\text{\tiny AT}}$\\
\hline
AMD
& 0.963& 0.987& 0.967& 0.990& 0.890& 0.937& 0.891& 0.936
\\
    \hline
    BLR
& \cellcolor{blue!25}0.989&\cellcolor{blue!25} 1.0& 0.976& 0.992& 0.973& \cellcolor{blue!25}0.998& 0.974& 0.991
\\
    \hline
    CHN
& 0.952& 0.959& 0.971& 0.972& 0.983& 0.985& 0.953& 0.969
\\
    \hline
    DEL
& 0.887& 0.866& \cellcolor{blue!25}1.0& \cellcolor{blue!25}1.0& 0.839& 0.850& \cellcolor{blue!25}1.0& \cellcolor{blue!25}0.998
\\
    \hline
    HYD
& 0.955& 0.983& 0.955& 0.985& 0.978& 0.967& 0.972& 0.977
\\
    \hline
    KOL
& 0.972& 0.914& 0.971& 0.922& 0.995& 0.943& 0.993& 0.942
\\
    \hline
    MUM
& 0.958& 0.984& 0.956& 0.990&\cellcolor{blue!25}1.0& 0.988&0.975&0.986
\\
   \hline
    AGT
& 0.916& 0.921& 0.942& 0.935& 0.944&0.939&0.946&0.941
\\
        \hline
    ASN
& \cellcolor{red!25}0.855& 0.930& \cellcolor{red!25}0.858& 0.933&0.839& 0.951&0.848&0.916
\\
        \hline
    BDI
& 0.902& 0.925& 0.936& 0.972&0.914& 0.952&0.943&0.928
\\
        \hline
        BGP
& 0.934& 0.910& 0.961& 0.962&0.968& 0.972&0.905&0.914
\\
        \hline
 BPL
& 0.933& 0.925& 0.944& 0.961& 0.974& 0.988& 0.940&0.965
\\ \hline
BSR
& 0.918& \cellcolor{red!25}0.861& 0.917& \cellcolor{red!25}0.889&0.913& 0.892&0.847&0.766
\\
        \hline
    BKN
& 0.896& 0.894& 0.942& 0.972&0.899& 0.924&0.941&0.961
\\
        \hline
    GWH
& 0.923& 0.902& 0.933& 0.921&0.853& 0.872&\cellcolor{red!25}0.725&\cellcolor{red!25}0.645
\\
        \hline
    IND
& 0.911& 0.956& 0.913& 0.976&0.896& 0.991&0.901&0.978
\\
        \hline
    KOP
& 0.879& 0.971& 0.891& 0.978&\cellcolor{red!25}0.577& 0.828&0.873&0.973
\\
        \hline
    LKO
& 0.955& 0.967& 0.949& 0.975&0.958& 0.987&0.913&0.949
\\
        \hline
    MAN
& 0.972& 0.990& 0.979& 0.986&0.938& 0.959&0.885&0.936
\\
        \hline
    PDY
& 0.964& 0.954& 0.975& 0.984&0.986& 0.989&0.986&0.936
\\
        \hline
    RPR
& 0.934& 0.970& 0.946& 0.981&0.910& 0.939&0.887&0.943
\\
        \hline
    SXR
& 0.985& 0.969& 0.981& 0.965&0.956& 0.950&0.948&0.923
\\
        \hline
    TVM
& 0.956& 0.930& 0.968& 0.939&0.878& \cellcolor{red!25}0.814&0.961&0.943
\\
        \hline
 VTZ
& 0.944& 0.919& 0.954& 0.940& 0.924& 0.918& 0.949&0.931
\\ \hline        
    \end{tabular}
    \vspace{0.2 cm}
    \caption{Relative conditional entropy ${\cal H}^R_{X,Y}$ for all the pollutant-weather variable pairs across cities. The maximum and minimum values for each pair are highlighted in blue and red, respectively.}
    \label{tab:relative}
\end{table*}
The values of the conditional entropy, ${\cal H}$, i.e., the uncertainty left in each pollutant after conditioning on RH or AT, computed from the aggregated data across all the chosen cities using Eq.~\eqref{eqn:8} are tabulated in Table~\ref{tab:Cond_Entropy} [see in Appendix]. 
The relative conditional entropy ${\cal H}^{R}$ (i.e, fraction of uncertainty remaining) in Table~\ref{tab:relative} is also calculated using the aggregated data from all selected cities using Eq.~\eqref{eqn:9}.
High ${\cal H}^{R}$ values close to 1 (shown in blue) indicate that meteorological factors explain very little of the pollutant variability.
In contrast, low ${\cal H}^{R}$ values (shown in red) suggest that RH and AT account for a significant portion of the variability in pollutant levels. In summary, smaller ${\cal H}^{R}$ values imply that meteorological parameters are highly informative about pollutant concentrations, whereas higher or near-unity values indicate that pollutant variations are largely independent of RH and AT.

The three metrics discussed in this section capture  various aspects of the pollutant-weather interactions. A consolidated summary of all dependence measures across locations is provided in Table~\ref{summary3}.
A more comprehensive understanding requires integrating these facets, the results of which we discuss in the next subsection. 

\subsection{Multi-Metric Compound Scoring of Pollutant–Meteorology Correlations}

This section  presents our key findings of a PCA-based composite assessment of the dependence of pollutants and meteorology variables derived from multiple correlation measures. We first examine intra-city patterns, focusing on the strength of dependence across all pollutant–meteorological variable pairs within individual cities. The analysis is then extended to a cross-city comparison, evaluating variability in composite correlation strength across all variable pairs among different urban locations. Based on the compound correlation score we then classify the cities into different clusters of varying pollutant-weather variables interaction strengths. 


\subsubsection{Intra-City Correlation Strength for each Pollutant-Weather Variable Pair}

At the first level of analysis, we use principal component analysis to quantify  the overall strength and structure of correlation between air pollutants and meteorological variables within each city. For a given city, we compute correlation strengths for all the pollutant–weather variable pairs using three distinct statistical metrics\footnote{It is to be noted that ignoring the directionality, we consider only the strength of the linear association of the correlation. Also note that, larger $\Tilde{\cal H}^R$ indicates weaker interaction and hence taking $h^R=(1-\Tilde{\cal H}^R)$ reverses its direction so that all three terms are on the same scale.}, namely, $|r_{X,Y}|$, $I_{X,Y}$ and $h^R_{X,Y}$.  
Thus, we have a city-specific data matrix of dimension  $8 \times 3,$  where each row corresponds to a particular pollutant–meteorological variable pair, and each column represents one correlation measure. 

PCA is applied to  the Z-score standardized values of this input matrix to extract the weights $w_k$ for the three features, i.e., the three correlation measures. These weights are then used to construct the first principal component [see Eq.~\eqref{eq:PC1}]. Thus, for the pollutant-weather variable pair $j=(X,Y)$, we have, 
\begin{align} \label{eqn:12}
    \tilde \lambda_{j=(X,Y)} = w_1 \, |\tilde{r}_{X,Y}| + w_2 \, \tilde{I}_{X,Y} + w_3 \,  \tilde{h}^R_{X;Y},
\end{align}
where $|\tilde{r}_{X,Y}|$, $\tilde{I}_{X,Y}$, and $\tilde{h}^R_{X;Y}$ denote the Z-score normalized versions of $|{r}_{X,Y}|$, ${I}_{X,Y}$, and ${h}^R_{X;Y}$, respectively. To convert it into a suitable composite correlation score, we shift the values to a non-negative scale by subtracting the minimum value from all the entries of the PC1 and then dividing by the range of $\tilde {\lambda}_j$. Thus, we arrive at a score,
  \begin{align} \label{eqn:13}
      \lambda_j = \frac{\tilde \lambda_j - \min(\tilde \lambda_j)}{\max(\tilde \lambda_j) - \min(\tilde \lambda_j)}
  \end{align}
This transformation preserves the relative ranking and can be interpreted as a strength-only measure which is bounded in the range $[0,1]$.

This exercise is repeated for each city. Thus, for the $i$-th city, we have an array $\lambda^{(i)}_j$, which provides a measure for the overall strength of the $j$-th pollutant-weather variable pair. The min-max normalization implies that, $0 \le \lambda^{(i)}_j \le 1$; larger value of $\lambda^{(i)}_j$ implies stronger correlation between the $j$-th pair. We refer to $\lambda^{(i)}_j$ as local correlation score (LCS) for the $j$-th variable pair in the $i$-th city. For all the cities, the first principal component, in this case, accounts for at least 69\% of the total variance. 
\begin{table}[t]
\centering
\renewcommand{\arraystretch}{2} 
\resizebox{\columnwidth}{!}{%
\begin{tabular}{|l|p{4cm}|p{1cm}|}
\hline
\textbf{Variable Pair} & \textbf{Cities with Very High LCS} & \textbf{No. of Cities} \\
\hline
PM\textsubscript{2.5}-RH & MUM (0.930), AMD (0.859), HYD (0.855), KOP (0.739) & 4 \\
\hline
PM\textsubscript{2.5}-AT & CHN (1), DEL (1), KOL (1), AGT (1), ASN (1), BGP (1), BPL (1), GWH (1), VTZ (1), LKO (0.957), BDI (0.919) & 11 \\
\hline
PM\textsubscript{10}-RH & AMD (1), BLR (1), HYD (1), MUM (1), BKN (1), KOP (1), IND (0.989), BPL (0.789) & 8 \\
\hline
PM\textsubscript{10}-AT & GWH (0.974), AGT (0.966), KOL (0.953), CHN (0.942), ASN (0.880), BKN (0.860), LKO (0.802), BDI (0.771) & 8 \\
\hline
SO\textsubscript{2}-RH & LKO (0.856), GWH (0.823), IND (0.772), MAN (0.725) & 4 \\
\hline
SO\textsubscript{2}-AT & LKO (1), TVM (1), BDI (0.855) & 3 \\
\hline
NO\textsubscript{2}-RH & IND (1), MAN (1), RPR (1), BPL (0.942), KOP (0.933), BLR (0.793) & 6 \\
\hline
NO\textsubscript{2}-AT & BDI (1), BSR (1), PDY (1), SXR (1), BKN (0.729) & 5 \\
\hline
\end{tabular}
}
\vspace{0.2 cm}
\caption{Cities exhibiting strong pollutant--weather variable pair associations (LCS $> 0.7$), identified via PCA.}
\label{tab:high_lcs_pairs}
\end{table}
Using Eqs.~\eqref{eqn:10}, \eqref{eq:PC1}, \eqref{eqn:12} and \eqref{eqn:13}, we compute the values of local correlation score for inter city pollutant--variable pairs.
Table~\ref{tab:high_lcs_pairs} presents the cities for which the LCS values are large ($>0.7$) for a given variable pair.
This intra-city analysis highlights which specific pollutant–meteorological interactions exhibit the strongest and most consistent relationships within each urban environment. Such localized insights are essential for identifying the most influential meteorological drivers of pollutant behavior and can guide city-specific environmental monitoring, forecasting models, or intervention strategies.

\subsubsection{Comprehensive Correlation Score: Cross-City Aggregation} 

Building on the intra-city analysis, we next extend the investigation to a cross-city scale, aiming to integrate and compare the strength of various pollutant–meteorology interactions across urban environments.

In this level  of analysis, we evaluate the spatial consistency and variability of pollutant–weather interactions by aggregating correlation metrics for each variable pair across all cities. This allows us to assess the extent to which specific pollutant–weather relationships exhibit coherent patterns or divergence across various locations. For each of the pollution-weather variable pairs we compile the corresponding correlation coefficients from all $m=24$ cities, using $p=3$ distinct correlation measures ($|\tilde r_{X,Y}|$, $\tilde I_{X,Y}$ and $h^R_{X,Y})$. This yields a pair specific input matrix of size $24 \times 3$ where each row represents a city and each column corresponds to a correlation measure. Each column of the matrix is then standardized via Z-score normalization, followed by PCA. The first principal component loadings are then used to construct the raw spatial correlation score $\tilde{\phi}_i^{(j)}$ for each variable pair $j$ across all the cities. 

To convert this to a suitable non-negative composite score,  we shift each value by subtracting the minimum of the observed values of $\tilde \phi_i$. 
\begin{equation} \label{eqn:14}
    \phi_i= \tilde \phi_i - \min(\tilde \phi_i)
\end{equation}
This procedure is repeated for all the $j$ variable pairs ($\forall j=1,\cdots,8.$). Thus, for the $j$-th variable pair, we arrive at a spatial correlaton score (SCS), $\phi_{i}^{(j)}$,which measures the overall strength of correlation in $j$-th pollutant-weather variable pair across all the cities ($\forall i=1,\cdots,24.$). The SCS values corresponding to each city for all the variable pairs are presented in Table~\ref{C_prime} in Appendix.

In the next stage of our analysis, we amalgamate the full spectrum of pollutant–weather correlation behavior across cities as well as variable pairs into a single, unified measure. To achieve this, we construct an input matrix of size $24 \times 8 $, where each entry $\phi_i^{(j)}$ represents the SCS value for $i^{th}$ city and $j^{\text{th}}$ variable pair. Evidently $i=1,2,3....24 $ and $j=1,2,...8$. This matrix reflects the complete landscape of localized correlation strengths across all urban environments and variable combinations. After applying Z-score normalization to this matrix, we perform PCA once more and extract the first principal component, which accounts for about 59.42\% of total variability.
This leads to a total correlation score of the  $i^{th}$ city, $\tilde{\mathcal{C}}(i)$, which is the weighted linear combination of the corresponding SCS values. This is expressed as [see Eq.~\eqref{eq:PC1}], 
\begin{align} 
\tilde{\mathcal{C}}(i) = &\,\sum_{j=1}^8w_j\tilde{f}_j(i),   \label{CCI_intercity} 
\end{align}
where, $\tilde{f}_j$ corresponds to the standardized value of SCS for the $j^{\text{th}}$ variable pair, and $w_j$ refers to the corresponding weight. The numerical values of the PCA loadings  are found to be, 
\begin{align} 
 w_1 &= 0.408.., w_2= 0.310.., w_3 = 0.331.., \cr 
 w_4 &= 0.376.., w_5 = 0.352.., w_6= 0.318.., \cr 
 w_7 &= 0.358.., w_8 = 0.365... \label{eqn:16}
\end{align}
and  are comparable across all the variable pairs. This observation is consistent with a subsequent PCA-based sensitivity analysis [See Sec.~\ref{sec:sen} below]. 

Once again,  we shift the values of the first principal component,
\begin{align} \label{eqn:17}
    {\cal C}(i) = \tilde {\cal C}(i) - \min\{\tilde {\cal C}(i)\},
\end{align}
to convert it to a non-negative score while preserving the relative ranking.  We refer to ${\cal C}(i)$ as Comprehensive Correlation Score (CCS), which
serves as a measure of overall pollutant-weather variable correlation strength in the cities.

\begin{figure}[tbh]
    \centering
    \includegraphics[width=0.9\linewidth]{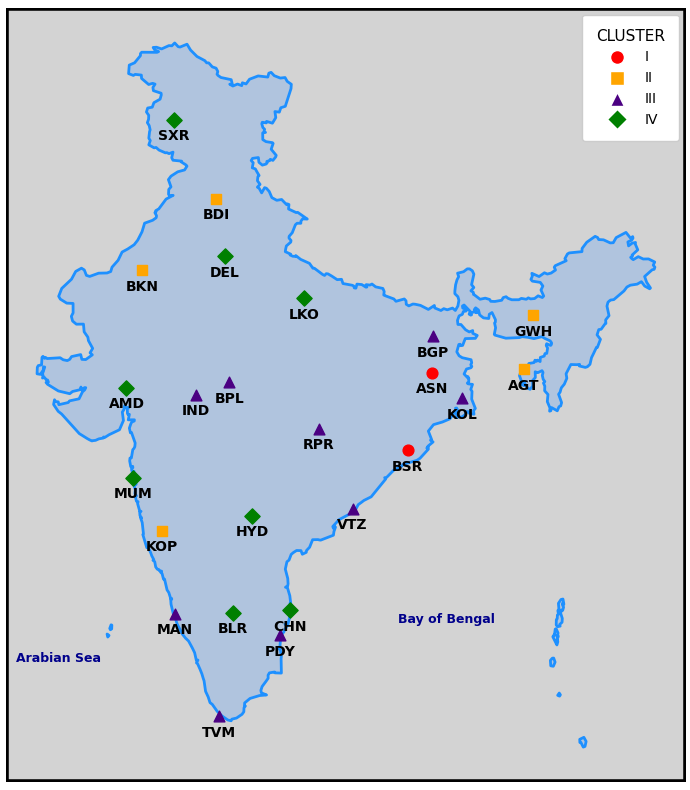}
    \caption{Spatial visualization of city clusters according to comprehensive correlation score $\mathcal{C}$. The map shows the cities corresponding to very high (red), high (orange), moderate (violet), and low (green) values of $\mathcal{C}$.}
    \label{fig:c_total_map}
\end{figure}

We apply k-means clustering to categorize cities into distinct groups based on the strength of their CCS ($\mathcal{C}$) values, revealing structural similarities in pollutant–weather interactions.
\begin{table}[t]
\renewcommand{\arraystretch}{1.4}
\begin{tabular}{|c|c|c||c|c|c|}
\hline     
\textbf{Cluster} & $\mathcal{C}$ & \textbf{Cities}&\textbf{Cluster} & $\mathcal{C}$ & \textbf{Cities}\\
\hline
 &  & & & 6.550 & BKN\\
 & 8.648 & ASN  & &  6.162 & GWH\\
 I&7.959&BSR&II & 5.714 &KOP \\
&&&& 5.642 &BDI \\
&&&& 4.761 & AGT\\
\hline
 & 4.403 & IND  &  & 2.142 & SXR\\
 & 4.375 &BPL &  & 2.011 & LKO \\
 & 4.019 & RPR &  & 1.696 & HYD\\
 & 3.632 &MAN & & 1.677 & MUM  \\
 III & 3.460 &BGP & IV  & 1.608 & DEL\\
 & 3.124 & KOL  &  & 1.554 & AMD\\
 & 2.651 & TVM &  & 0.959 & CHN\\
 & 2.650 & PDY &   & 0 & BLR\\
  & 2.636 &VTZ &&& \\
\hline
\end{tabular}
\vspace{0.2 cm}
\caption{Classification of cities based on Comprehensive Correlation Score ($\mathcal{C}$).}
\label{C_Total}
\end{table}
We compute comprehensive correlation score in the cities using Eq.~\eqref{eqn:17} and then based on these score, we apply k-means clustering to categorize them into different groups. This leads to four distinct clusters: 
\begin{itemize}
    \item Cluster I (Very high): $\mathcal{C} \geq 7.959$
    \item Cluster II (High): $4.761 \leq \mathcal{C} \leq 6.550$
    \item Cluster III (Moderate): $2.636 \leq \mathcal{C}\leq 4.403$
    \item Cluster IV (Low): $\mathcal{C} \leq 2.142$
\end{itemize}
The cities belonging to these clusters and their corresponding CCS values are summarized in Table~\ref{C_Total}.
This classification  which reflects systemic pollutant–weather interaction strength, provides a valuable basis for region-specific environmental modeling, comparative vulnerability assessments, and the design of transferable mitigation strategies tailored to cities with similar correlation dynamics. Fig.~\ref{fig:c_total_map} depicts  the geographic visualization of city clusters of similar correlation strength.  
We find that most megacities exhibit relatively low to moderate total correlation strength, reflecting complex and possibly more heterogeneous pollutant-meteorology interactions. In contrast, cities like Asansol and Bhubaneswar, which are comparatively less populated, show significantly higher correlations, indicating a stronger link between atmospheric conditions and pollutant fluctuations. These findings reveal important spatial variability in pollutant-meteorology dynamics across urban environments and provide a foundation for further theoretical exploration of pollutant-meteorology relationships across diverse spatial scales.

\subsubsection{Sensitivity Analysis}\label{sec:sen}
Sensitivity analysis is used to assess the extent to which individual variable-pairs contribute to the comprehensive correlation score $\mathcal{C}$. We have noted that, in the computation of CCS, the first principal component accounts for the maximum variability $\Sigma=59.42\%$ , explained by all the variable pairs. Let, the percentage of reduction in variance ($\Delta \Sigma_j$) due to exclusion of  the \(j^{\text{th}}\) variable pair be given by,
\begin{equation} \label{eqn:18}
\Delta \Sigma_j (\%) = \frac{\Sigma - \Sigma_j}{\Sigma}\times 100,
\end{equation}
where, $\Sigma_{j}$  denotes the variance accounted for by the first principal component when the $j^{\text{th}}$ variable pair is excluded and PCA is reapplied  to  the remaining $j-1$ variable pairs.
In practice, small values of \(|\Delta \Sigma_j|\), i.e., $\leq 3\%$ are considered negligible, indicating that the PCA structure is robust to the removal of the corresponding variable pair. On the other hand, values exceeding $5\%$ are typically regarded as strong, indicating that the excluded variable pair has a significant effect on the PC1 variance.

\begin{table}[th]
\centering
\resizebox{0.8\columnwidth}{!}{%
\begin{tabular}{|p{1.8 cm}|p{1.2cm}|p{1.2cm}|}\hline
Excluded pair  & \textbf{$\Sigma_j$ (\%)} & \textbf{$\Delta \Sigma_j (\%)$} \\\hline
PM\textsubscript{2.5}–RH & 57.25 & -2.17 \\\hline
PM\textsubscript{2.5}–AT & 62.36 & 2.94 \\\hline
PM\textsubscript{10}–RH & 61.47 & 2.05 \\\hline
PM\textsubscript{10}–AT & 59.26 & -0.16 \\\hline
SO\textsubscript{2}–RH & 60.43 & 1.01 \\\hline
SO\textsubscript{2}–AT & 61.92 & 2.50 \\\hline
NO\textsubscript{2}–RH & 60.23 & 0.80 \\\hline
NO\textsubscript{2}–AT & 59.86 & 0.44 \\\hline
\end{tabular}}
\caption{Change in percentage of variance explained for the first principal component upon individual exclusion of each variable pair.
}
\label{sensitivity}
\end{table}
Table~\ref{sensitivity} presents the results of the sensitivity analysis, which we compute using Eq.~\eqref{eqn:18}, showing the variance explained by the first principal component after sequentially excluding each of the eight pollutant–meteorology variable pairs and reapplying PCA on the remaining set. The table also includes the percentage change in explained variance resulting from the exclusion of each pair. The results indicate that  
the variance loss from removing any one pair is minimal, suggesting that the comprehensive correlation score is robust and not overly dependent on any specific interaction. This further reinforces the consistency of the combined effect across multiple cities. However,  amongst all pairs,  PM\textsubscript{2.5}-RH pair appears to be the most influential, with its removal causing a modest decrease of  variance (2.17\%) whereas interaction pairs like PM\textsubscript{2.5}-AT, SO\textsubscript{2}-AT and PM\textsubscript{10}-RH record a slight increase (almost 2.50\%) in explained variance when excluded.
\subsection{Temporal Correlation}

This subsection presents the results of the temporal correlation analysis between pollutants and meteorological variables. Using time-domain measures, we examine the directionality and lead–lag structure of their interactions, highlighting key temporal dependencies that extend beyond static associations.

\subsubsection{Transfer Entropy}
\begin{table*}[th]
\small
\centering
\begin{tabular}{|c|p{1.6cm}|p{1.55cm}|p{1.5cm}|p{1.5cm}|p{1.35cm}|p{1.33cm}|p{1.33cm}|p{1.33cm}|}
\hline 
City & $\Delta {\cal T}_{(\text{\tiny PM}_{2.5},\text{\tiny RH})}$ & $\Delta {\cal T}_{(\text{\tiny PM}_{2.5},\text{\tiny AT})}$ & $\Delta {\cal T}_{(\text{\tiny PM}_{10},\text{\tiny RH})}$ & $\Delta {\cal T}_{(\text{\tiny PM}_{10},\text{\tiny AT})}$ & $\Delta {\cal T}_{(\text{\tiny NO}_{2},\text{\tiny RH})}$ & $\Delta {\cal T}_{(\text{\tiny NO}_{2},\text{\tiny AT})}$ & $\Delta {\cal T}_{(\text{\tiny SO}_{2},\text{\tiny RH})}$ & $\Delta {\cal T}_{(\text{\tiny SO}_{2},\text{\tiny AT})}$ \\
\hline
AMD & \cellcolor{red!25}-0.035& \cellcolor{blue!25}0.006&\cellcolor{red!25}-0.031&\cellcolor{blue!25}0.017 &\cellcolor{red!25}-0.033 & \cellcolor{blue!25}0.023&\cellcolor{red!25}-0.025 &\cellcolor{blue!25}0.002\\
\hline
BLR  &\cellcolor{red!25}-0.007&\cellcolor{red!25}-0.004&\cellcolor{red!25}-0.007&\cellcolor{blue!25}0.020&\cellcolor{red!25}-0.017&\cellcolor{blue!25}0.007&\cellcolor{red!25}-0.007&\cellcolor{blue!25}0.011\\
\hline
CHN  &\cellcolor{red!25}-0.005&\cellcolor{blue!25}0.009&\cellcolor{red!25}-0.002&\cellcolor{red!25}-0.001&\cellcolor{red!25}-0.003&\cellcolor{red!25}-0.003&\cellcolor{red!25}-0.011&\cellcolor{blue!25}0.002\\
\hline
DEL  &\cellcolor{blue!25}0.008&\cellcolor{blue!25}0.027&\cellcolor{blue!25}0.008&\cellcolor{blue!25}0.026&\cellcolor{red!25}-0.001&\cellcolor{blue!25}0.005&\cellcolor{blue!25}0.002&\cellcolor{blue!25}0.006\\
\hline
HYD  &\cellcolor{blue!25}0.064&\cellcolor{blue!25}0.071&\cellcolor{blue!25}0.066&\cellcolor{blue!25}0.075&\cellcolor{red!25}-0.002&\cellcolor{blue!25}0.009&\cellcolor{blue!25}0.005&\cellcolor{blue!25}0.008\\
\hline
KOL  &\cellcolor{red!25}-0.021&\cellcolor{blue!25}0.018&\cellcolor{red!25}-0.025&\cellcolor{blue!25}0.008&\cellcolor{red!25}-0.019&\cellcolor{blue!25}0.013&\cellcolor{blue!25}0.001&\cellcolor{blue!25}0.001\\
\hline
MUM &\cellcolor{red!25}-0.014&\cellcolor{blue!25}0.003&\cellcolor{red!25}-0.008&\cellcolor{blue!25}0.005&\cellcolor{red!25}-0.004&\cellcolor{blue!25}0.005&\cellcolor{red!25}-0.003&\cellcolor{blue!25}0.003\\
\hline
AGT &\cellcolor{red!25}-0.050&\cellcolor{blue!25}0.060& \cellcolor{red!25}-0.035&\cellcolor{blue!25}0.055&\cellcolor{blue!25}0.017&\cellcolor{blue!25}0.146&\cellcolor{red!25}-0.130&\cellcolor{red!25}-0.052\\
\hline
ASN  &\cellcolor{red!25}-0.078&\cellcolor{blue!25}0.199&\cellcolor{red!25}-0.163&\cellcolor{blue!25}0.099&\cellcolor{red!25}-0.154&\cellcolor{blue!25}0.096&\cellcolor{red!25}-0.011&\cellcolor{blue!25}0.229\\
\hline
BDI  &\cellcolor{red!25}-0.007&\cellcolor{blue!25}0.127&\cellcolor{blue!25}0.063&\cellcolor{blue!25}0.208&\cellcolor{blue!25}0.106&\cellcolor{blue!25}0.317&\cellcolor{blue!25}0.083&\cellcolor{blue!25}0.298\\
\hline
BGP  & \cellcolor{red!25}-0.023&\cellcolor{blue!25}0.067&\cellcolor{blue!25}0.022&\cellcolor{blue!25}0.098&\cellcolor{red!25}-0.103&\cellcolor{red!25}-0.015&\cellcolor{red!25}-0.039&\cellcolor{blue!25}0.018\\
\hline
BPL  &\cellcolor{red!25}-0.036&\cellcolor{blue!25}0.067&\cellcolor{blue!25}0.037&\cellcolor{blue!25}0.112&\cellcolor{blue!25}0.032&\cellcolor{blue!25}0.129&\cellcolor{red!25}-0.054&\cellcolor{blue!25}0.057\\
\hline
BSR  &\cellcolor{blue!25}0.025&\cellcolor{blue!25}0.117&\cellcolor{blue!25}0.055&\cellcolor{blue!25}0.226&\cellcolor{red!25}-0.073&\cellcolor{blue!25}0.055&\cellcolor{red!25}-0.063&\cellcolor{red!25}-0.039\\
\hline
BKN  &\cellcolor{red!25}-0.104&\cellcolor{blue!25}0.112&\cellcolor{blue!25}0.067&\cellcolor{blue!25}0.383&\cellcolor{red!25}-0.032&\cellcolor{blue!25}0.273&\cellcolor{red!25}-0.084&\cellcolor{blue!25}0.168\\
\hline
GWH  &\cellcolor{blue!25}0.018&\cellcolor{blue!25}0.044&\cellcolor{blue!25}0.040&\cellcolor{blue!25}0.108&\cellcolor{red!25}-0.001&\cellcolor{blue!25}0.048&\cellcolor{red!25}-0.056&\cellcolor{red!25}-0.023\\
\hline
IND &\cellcolor{red!25}-0.032&\cellcolor{blue!25}0.069&\cellcolor{blue!25}0.125&\cellcolor{blue!25}0.250&\cellcolor{blue!25}0.038&\cellcolor{blue!25}0.167&\cellcolor{blue!25}0.082&\cellcolor{blue!25}0.090\\
\hline
KOP &\cellcolor{red!25}-0.085&\cellcolor{blue!25}0.021&\cellcolor{red!25}-0.043&\cellcolor{blue!25}0.083&\cellcolor{red!25}-0.068&\cellcolor{blue!25}0.049&\cellcolor{blue!25}0.017&\cellcolor{blue!25}0.011\\
\hline
LKO  &\cellcolor{blue!25}0.016&\cellcolor{blue!25}0.070&\cellcolor{red!25}-0.005&\cellcolor{blue!25}0.025&\cellcolor{red!25}-0.010&\cellcolor{blue!25}0.034&\cellcolor{red!25}-0.042&\cellcolor{blue!25}0.040\\
\hline
MAN  &\cellcolor{blue!25}0.079&\cellcolor{blue!25}0.150&\cellcolor{blue!25}0.093&\cellcolor{blue!25}0.184&\cellcolor{blue!25}0.135&\cellcolor{blue!25}0.097&\cellcolor{blue!25}0.024&\cellcolor{blue!25}0.146\\
\hline
PDY &\cellcolor{red!25}-0.101&\cellcolor{blue!25}0.072&\cellcolor{red!25}-0.060&\cellcolor{blue!25}0.163&\cellcolor{red!25}-0.115&\cellcolor{blue!25}0.043&\cellcolor{red!25}-0.104&\cellcolor{blue!25}0.066\\
\hline
RPR &\cellcolor{red!25}-0.057&\cellcolor{blue!25}0.009&\cellcolor{blue!25}0.015&\cellcolor{blue!25}0.092&\cellcolor{red!25}-0.162&\cellcolor{red!25}-0.072&\cellcolor{red!25}-0.040&\cellcolor{red!25}-0.010\\
\hline
SXR  &\cellcolor{red!25}-0.272&\cellcolor{blue!25}0.026&\cellcolor{red!25}-0.247&\cellcolor{blue!25}0.045&\cellcolor{red!25}-0.356&\cellcolor{red!25}-0.007&\cellcolor{red!25}-0.171&\cellcolor{blue!25}0.041\\
\hline
TVM  &\cellcolor{blue!25}0.009&\cellcolor{blue!25}0.029&\cellcolor{blue!25}0.003&\cellcolor{blue!25}0.002&\cellcolor{blue!25}0.003&\cellcolor{red!25}-0.030&\cellcolor{red!25}-0.002&\cellcolor{blue!25}0.026\\
\hline
VTZ &\cellcolor{blue!25}0.048&\cellcolor{blue!25}0.057&\cellcolor{blue!25}0.058&\cellcolor{blue!25}0.118&\cellcolor{blue!25}0.088&\cellcolor{blue!25}0.141&\cellcolor{blue!25}0.004&\cellcolor{blue!25}0.018\\
\hline
\end{tabular}
\vspace{0.2cm}
\caption{Spatial distribution of directional information flow ($\Delta {\cal T }$) for pollutant–weather variable pairs. Positive (blue)  and negative (red) values are color-coded to indicate the dominant direction of information flow.}
\label{tab:TE1}
\end{table*}
We compute the transfer entropies, ${\cal T}_{X\to Y}$ and ${\cal T}_{Y\to X}$, using Eq.~\eqref{eq:Tyx_2} for all pollutant-weather pairs $(X,Y)$ in each city. To quantify the dominant direction of information flow, we define $\Delta {\cal T}_{X,Y} = [{\cal T}_{X\to Y}\,-\,\,{\cal T}_{Y\to X}]$.
A positive value ($\Delta {\cal T}_{X,Y}>0$) implies that pollutants ($X$) leads meteorological parameters ($Y$), whereas a negative value ($\Delta {\cal T}<0$) indicates that $Y$ leads $X$.
As seen from Table~\ref{tab:TE1}, $\Delta {\cal T}_{X,RH}<0$ for the majority of cities, suggesting that relative humidity (RH) predominantly drives variations in pollutant concentrations. However, a smaller subset of cities, including Delhi, Hyderabad, Bhubaneswar, Guwahati, Lucknow, Mangaluru, Thiruvananthapuram, and Visakhapatnam, exhibits the opposite pattern, where PM\textsubscript{2.5} changes precede those in RH. Similar trends are observed for other pollutant–RH pairs such as ($\text{PM}_{10}$, RH), ($\text{NO}_{2}$, RH), and ($\text{SO}_{2}$, RH), where most cities show RH as the leading variable, though a few instances with $\Delta {\cal T}_{X,RH}>0$ also occur, as reported in Table~\ref{tab:TE1}.
Conversely, nearly all cities show $\Delta {\cal T}_{X,AT}>0$ for pollutant–ambient temperature pairs, indicating that pollutant variations generally precede those in AT. Only a few exceptions with $\Delta {\cal T}_{X,AT}<0$ are observed. Thus, in most of the locations where RH leads the pollutants, AT tends to lag behind them.

It is noteworthy that the magnitudes of $\Delta {\cal T}$ remain close to zero across most locations, with the sole exception of Srinagar, Baddi, Asansol and Bikaner likely reflecting its distinct meteorological regime due to geographic factors. All other values are substantially smaller, indicating limited directional asymmetry in information flow. The relatively small values of $\Delta {\cal T}$, irrespective of the correlation strength between pollutants and weather variables, suggest that neither variable solely governs the underlying dynamics. Instead, the comparable TE magnitudes in both directions across cities indicate a bidirectional relationship between pollutants and meteorological parameters characterized by their mutual influence~\cite{Chen_2017}. This highlights the importance of considering these variable pairs as components of a coupled dynamical system in the study of air quality.

\subsubsection{Time-Delayed Mutual Information}

We evaluate $I_{X,Y}(\tau)$ using Eq.~\eqref{eqn:23} for $\tau \in [-10,10]$ days, with $X$ as all four pollutants and $Y$ as the two weather variables. 
We find that, for particulate matter-RH pairs [($\text{PM}{_{2.5}}$, RH) and ($\text{PM}{_{10}}$, RH)], the TDMI curves exhibit a global maximum at zero lag ($\tau = 0$) and subsequently decays to constant values for large positive and negative lags. This behaviour is shown in Fig.~\ref{fig:TDMI_extra} (in Appendix)  for a few representative cities since most locations exhibit similar behavior for the particulate matter–RH pairs.
A comparable trend is also observed for the gaseous pollutant-RH pairs [($\text{NO}{_{2}}$, RH) and ($\text{SO}{_{2}}$, RH)], though the number of cities displaying this pattern is smaller.
In contrast, no consistent or discernible pattern emerges for the pollutant–AT (ambient temperature) pairs. While we analyzed these cases as well, the absence of meaningful structure in their TDMI profiles led us to exclude them from further discussions.

The TDMI analysis for pollutant-RH pairs provides valuable insight into how mutual information evolves over time, revealing the characteristic response timescale of pollutant dynamics to humidity variations.  
As shown in Fig.~\ref{fig:TDMI_extra} (in Appendix), the TDMI decays rapidly and displays an exponential dependence for $\tau>0$. Accordingly, we propose the following form for the decay,
\begin{equation}
\Delta_{X,Y}(\tau) = I_{X,Y}(\tau) - I_{X,Y}^0 \approx ae^{-\tau/\tau_0},
\end{equation}
which provides a good fit across all pollutant–RH pairs $(X,Y)$.
Here,  $I_{X,Y}^0$ denotes the saturation value at large positive $\tau$, $a$ is a constant prefactor and $\tau_0$ is the typical time-scale of decay. These parameters depend on the city and the variable pair considered.

\begin{figure*}[tbh]
    \centering    \includegraphics[width=0.49\linewidth]{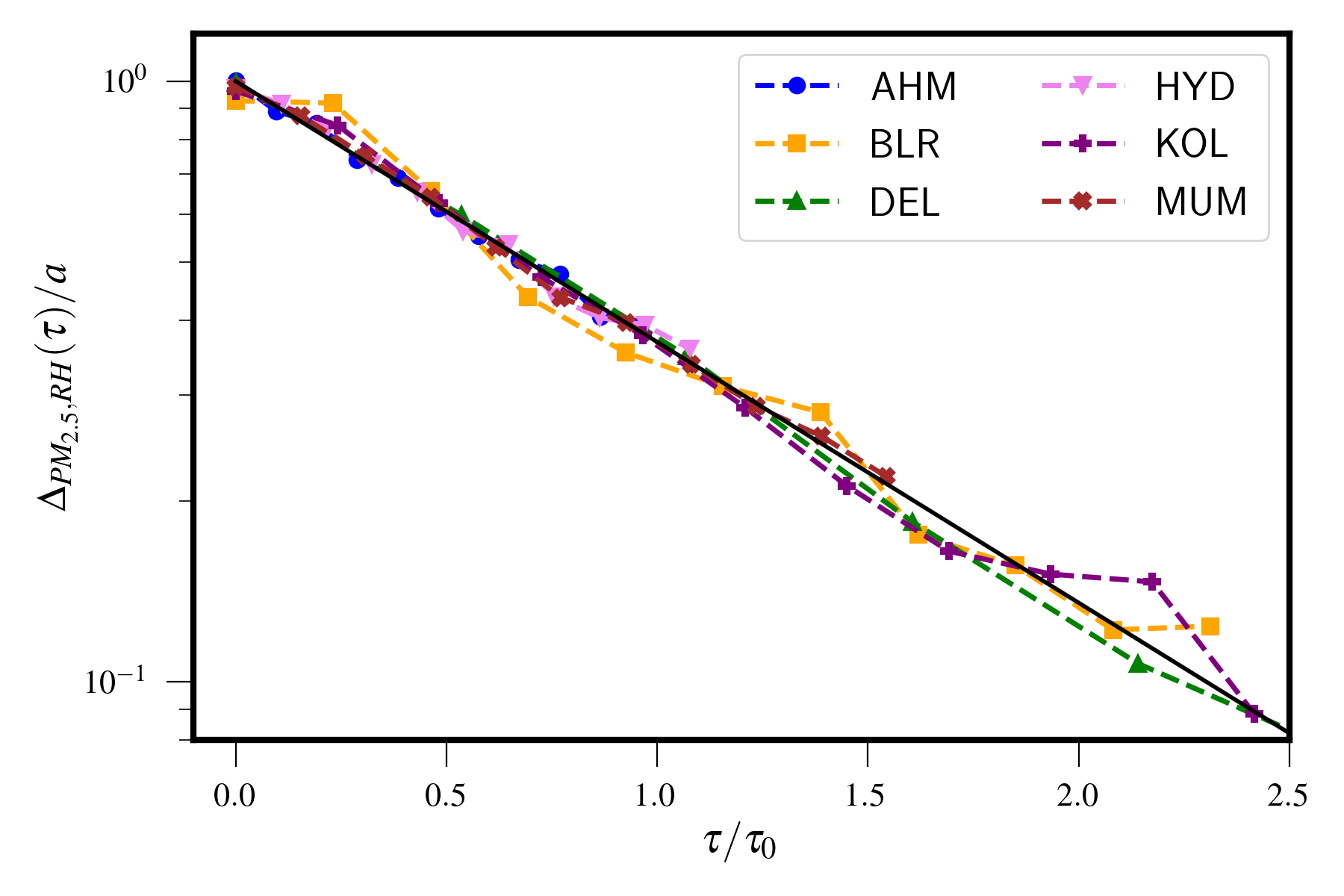}  \includegraphics[width=0.49\linewidth]{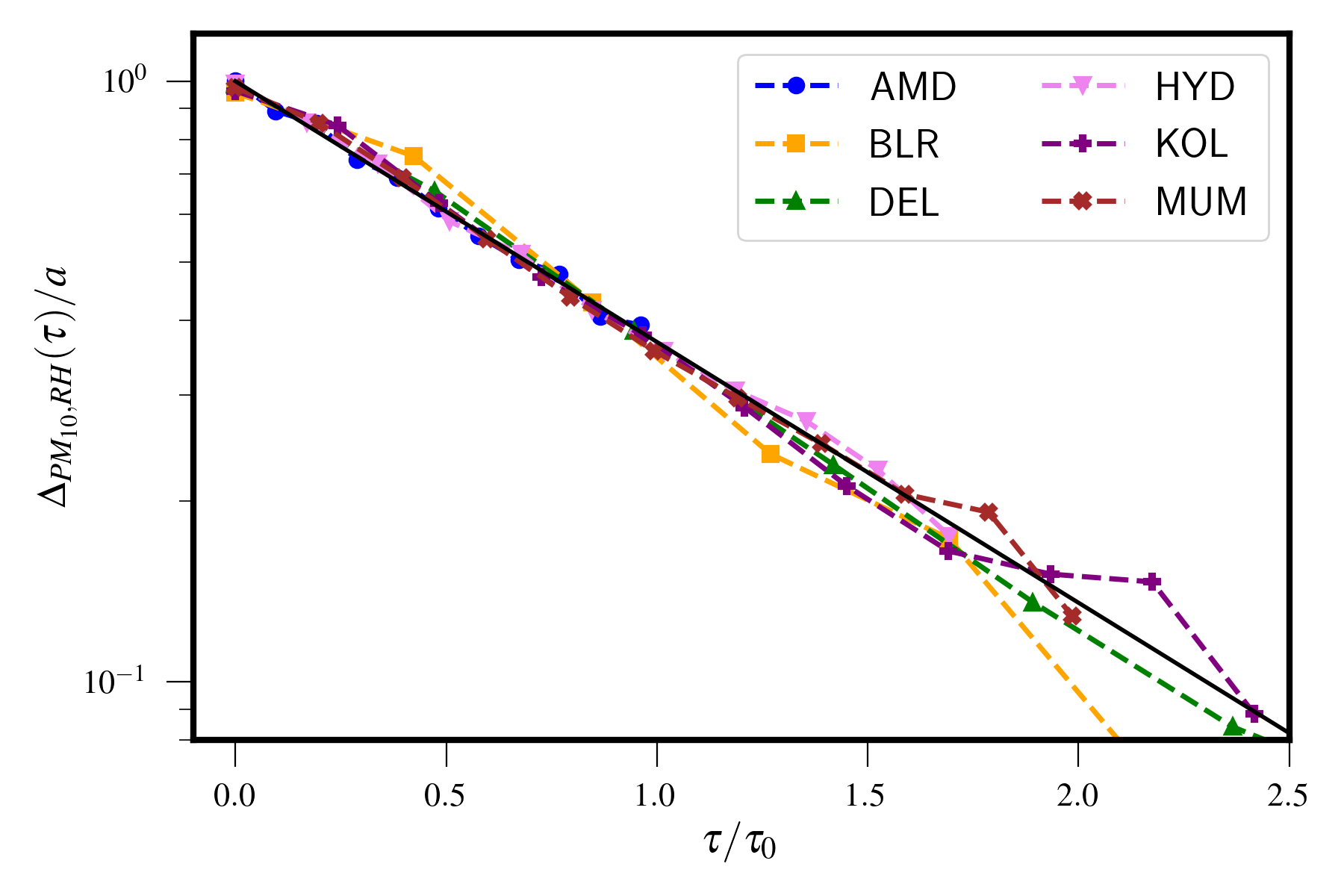}   
\includegraphics[width=0.49\linewidth]{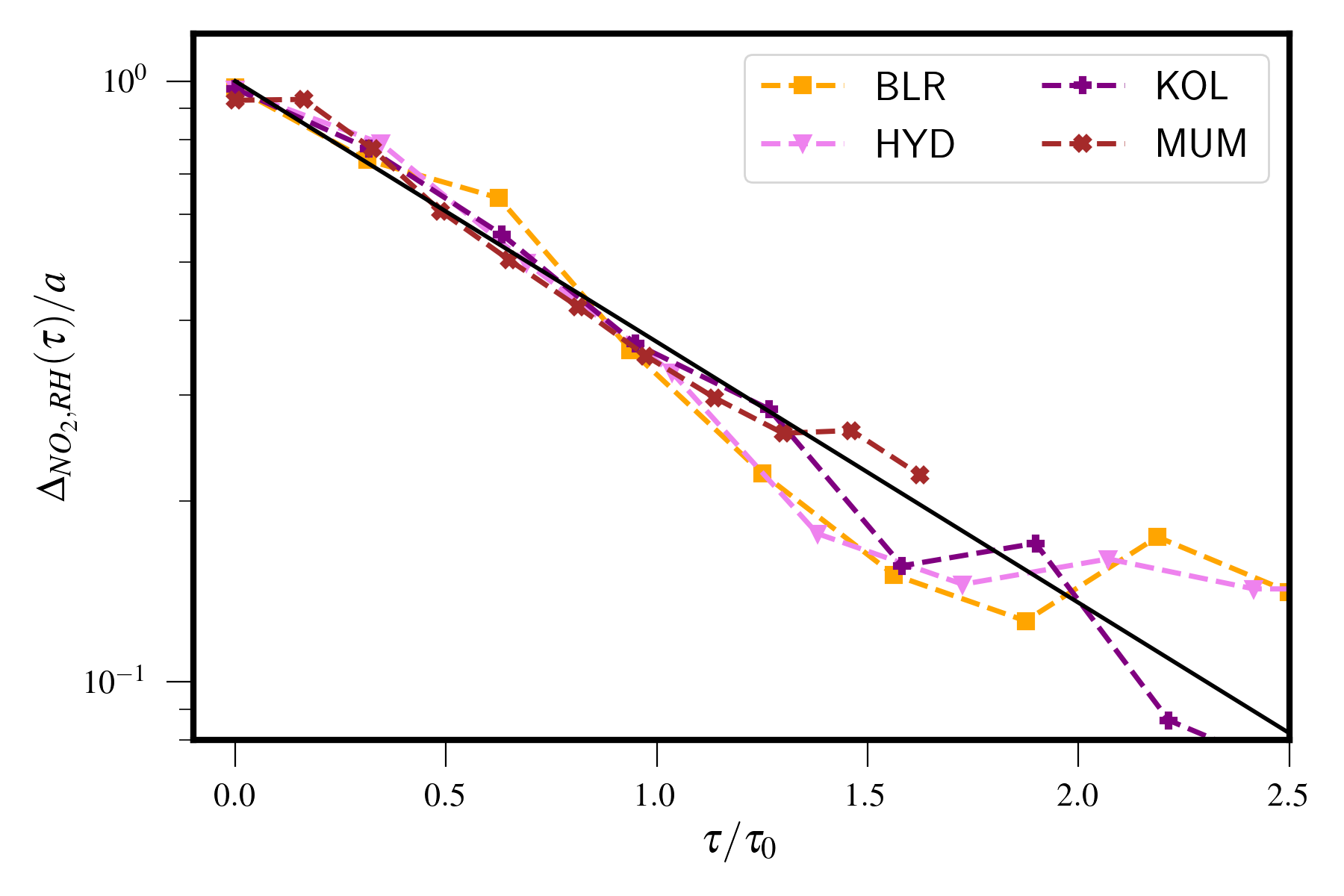} \includegraphics[width=0.49\linewidth]{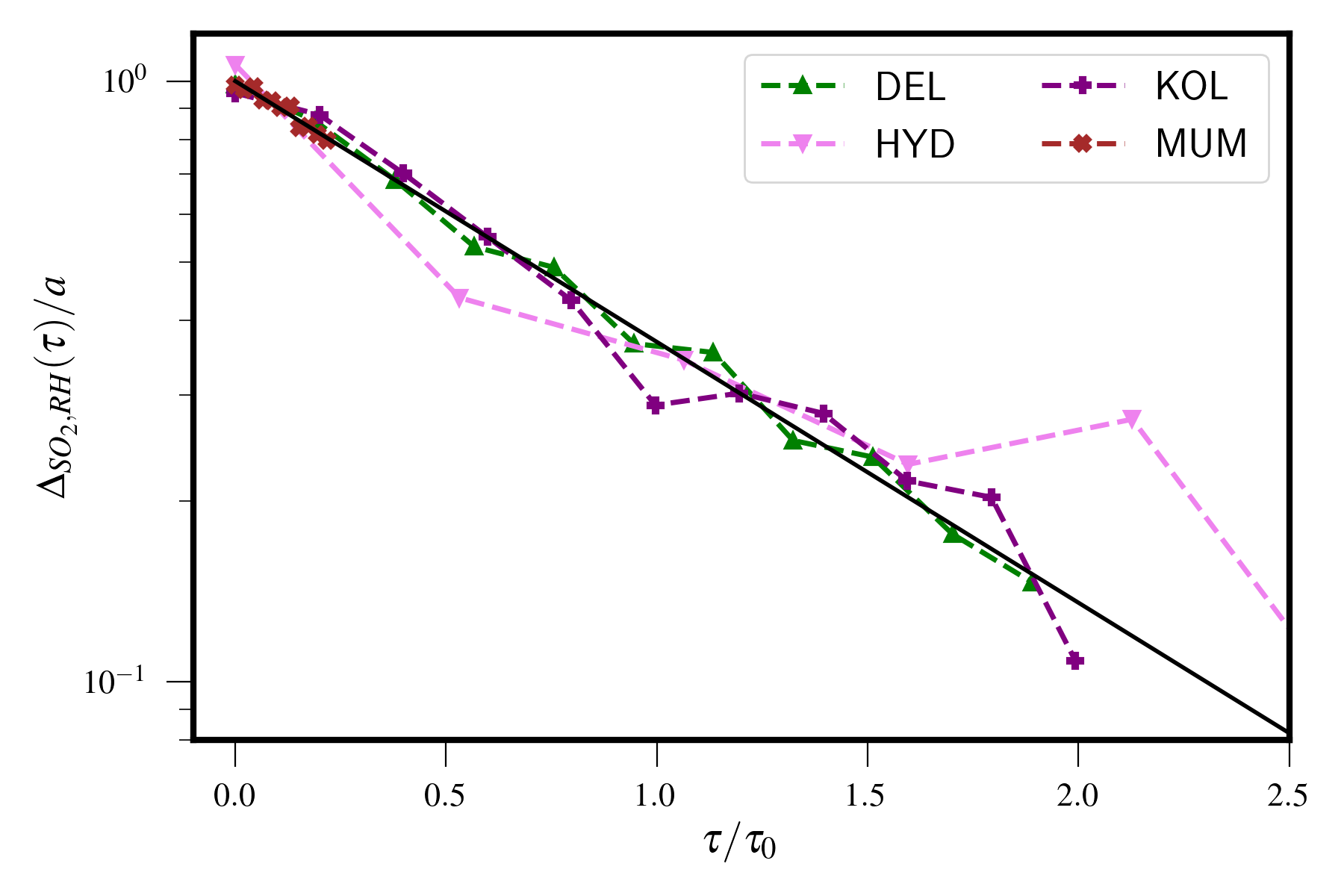}
\caption{Exponential decay for all four pollutant-RH pairs across selected representative cities, plotted on a semi-logarithmic scale with $\Delta_{X,Y}/a$ as a function of the scaled variable $\tau/\tau_0$. In each panel, the black solid line denotes the function $e^{-\tau/\tau_0}$. 
}
    \label{fig:TDMI_low}
\end{figure*}

This exponential decay is illustrated in Fig.~\ref{fig:TDMI_low} where we present a semilog plot of $\Delta_{X,Y}(\tau)/a$ as a function of the scaled variable $\tau/\tau_0$ for a set of selected cities for all pollutant-RH pair. 
This behavior suggests an almost instantaneous adjustment of pollutant concentrations to humidity fluctuations, with the influence of any individual RH perturbation dissipating rapidly—typically within a few days. Such rapid decay indicates limited temporal persistence in pollutant–RH interactions.
However, certain cities display TDMI profiles that deviate markedly from this general trend (which are not shown here).
In these cases, MI peaks 
at a nonzero lag (either positive or negative), and the decay deviates 
noticeably from an exponential form. These features suggest a delayed and more persistent response of fine particulate levels to humidity changes, likely arising from distinct local meteorological or emission conditions.

\section{Discussion}\label{sec:disc}

Meteorological variables exhibit heterogeneous influences on pollutant concentrations with variation in strength and pollutant types across urban regions.
While earlier works~\cite{gupta_2018,a,b,c} primarily relied on individual correlation measures, the PCA-based composite index used here provides a robust integration of multiple metrics enabling systematic comparison across locations. 
This index also supports the classification of urban regions according to the strength of pollutant--meteorology dependence, highlighting distinct groups of cities with relatively strong or weak composite correlations and offering a structured perspective on spatial heterogeneity.
Regions with stronger composite dependence are likely dominated by meteorological controls on dispersion, transformation, or accumulation, whereas weaker dependence may indicate a greater influence of local emission patterns. 

The unified correlation strength, transfer entropy, and time-delayed mutual information provide a physically coherent picture of pollutant–meteorology interactions. The composite correlation score developed here identifies where meteorological control is strong, while TE reveals the direction of influence and TDMI characterises the timescales of response. The observed spatial and temporal patterns are consistent with known atmospheric processes governing dispersion, transport, and chemical transformation, lending physical credibility to the information-theoretic framework.


The PCA-based integration of multiple correlation measures advances beyond single-metric approaches. It reduces metric-specific sensitivity and improves interpretability across large multi-city pollutant–meteorology datasets.
The resulting classification translates statistical dependence patterns into meaningful spatial groupings, with potential applications in air quality management and comparative assessments.

Several limitations warrant consideration. Using daily-averaged data may blur  short-term processes such as diurnal cycles or  transient pollution events. The focus on pairwise dependencies does not capture multivariate interactions or nonlinear feedbacks among pollutants and meteorological variables, and seasonal variability is not explicitly addressed. The PCA-based approach, while useful for integration, can lose information from lower-variance but potentially important structures and may under-represent nonlinear behavior. These considerations suggest that PCA-based composites should complement, rather than replace, more detailed process-level analyses.

Despite these limitations, the framework has important implications. The classification of urban regions based on composite dependence strength can inform region-specific air quality management strategies, support comparative assessments across cities, and provide a foundation for integrating meteorological sensitivity into predictive or scenario-based modeling. Overall, the PCA-based composite assessment offers a robust and transferable tool for understanding pollutant--meteorology relationships, advancing both methodological practice and interpretive understanding in multi-city air quality studies.

\section{Conclusion}\label{sec:concl}

Our work contributes to the expanding application of statistical physics and information-theoretic tools to the study of real-world environmental systems.
Using entropy-based measures, we analyze both static interdependence and dynamic information flow to investigate correlations, fluctuations, and variability between air pollutants and meteorological variables across multiple Indian cities. 
By combining both linear and nonlinear dependency measures, namely, Pearson correlation, mutual information, relative conditional entropy, as well as  transfer entropy and time-delayed mutual information, we build a comprehensive picture of how 
environmental variables jointly evolve in urban atmospheres. 
In the static analysis, to aggregate these diverse correlation patterns, we introduce a multi-level compound measure of pollutant–meteorological correlation strength. Using principal component analysis, we first quantify the intra-city correlation strength (LCS) for each pollutant–weather variable pair, identifying the most dominant interactions within each urban environment. This localized assessment provides valuable insight into the specific meteorological drivers governing pollutant variability, supporting the design of city-specific air quality management and forecasting models.
Extending the framework to a cross-city scale, we derive a comprehensive correlation score ($\mathcal{C}$) that enables comparison and ranking of systematic  pollutant–meteorology interaction strengths across cities. 
This comparative amalgamation uncovers notable spatial heterogeneity. Most megacities exhibit relatively low to moderate overall correlations, reflecting complex and heterogeneous atmospheric processes  whereas several less populated cities demonstrate markedly stronger pollutant–weather coupling. These contrasts highlight the role of regional climatic regimes, emission characteristics, and urban morphology in shaping local air quality dynamics. Furthermore, PCA-based sensitivity analysis indicates that no single pollutant–weather variable pair exerts a dominant influence, highlighting the multifactorial nature of these interactions.
Overall, the proposed multi-metric and multi-scale framework not only advances understanding of pollutant–meteorology relationships but also establishes a quantitative basis for developing region-specific forecasting models and adaptive mitigation strategies.

We further study the temporal structure of pollutant-weather variable correlations through transfer entropy and time-delayed mutual information. TE analysis of the pollutant–weather variable pairs show that in most locations where RH leads the pollutants, AT tends to lag behind them, highlighting contrasting temporal roles of these meteorological drivers in influencing pollutant dynamics. Although the values of $\Delta {\mathcal{T}}$ are very small in several cities showing bidirectional information flow, suggesting feedback mechanisms between pollutant levels and meteorological parameters. The TDMI results further indicate that, in many cases, mutual information peaks at lag zero and decays exponentially with increasing lag, implying predominantly contemporaneous interactions with limited temporal memory, reflecting short-lived correlations. These decay profiles may serve as useful indicators of system responsiveness and reactivity.

More broadly, the integrative entropy-based approach demonstrated here provides a
promising foundation for future work incorporating seasonal regimes, multivariate
interactions, and nonlinear dynamical modeling. Such extensions may further
enhance understanding of urban atmospheric processes and support region-specific
mitigation and adaptation strategies.

To the best of our knowledge, this work represents the first attempt in this direction, advancing the understanding of spatiotemporal dependencies in urban air quality and supporting region-specific assessments and monitoring under data-scarce or noisy conditions.


\section{Acknowledgment}

The authors  sincerely thank the anonymous Associate Editor of the journal ``Stochastic Environmental Research and Risk Assessment'' for the thoughtful and constructive feedback provided on an earlier version of this manuscript. All these  comments greatly helped to improve the methodological rigor and overall quality of the study. The authors appreciate  the contribution of Moumita De in partial data collection during the initial phase of this work. KG acknowledges  the support provided by the Indian Statistical Institute, Kolkata during this project. 

\section{Statements and Declarations}
\begin{itemize}
 \item \textbf{Data availability Statement}: All the data used in this work are publicly available~\cite{1}.\\
   \item \textbf{Competing Interests:} There are no competing interests to declare.\\
   \item \textbf{Funding:} No funding information to declare. \\
   \item \textbf{Author contributions:} Koyena Ghosh contributed to data collection, data curation, analysis, visualization, methodology and writing of the draft. Suchismita Banerjee contributed to data curation, analysis, visualization, methodology and writing of the draft. Urna Basu and Banasri Basu contributed to conceptualization, development of the methodology, analysis, supervision and writing of the draft. 
 \end{itemize}
 
\section*{Appendix}

\begin{figure}
    \centering   
    \includegraphics[width=1.1\linewidth]{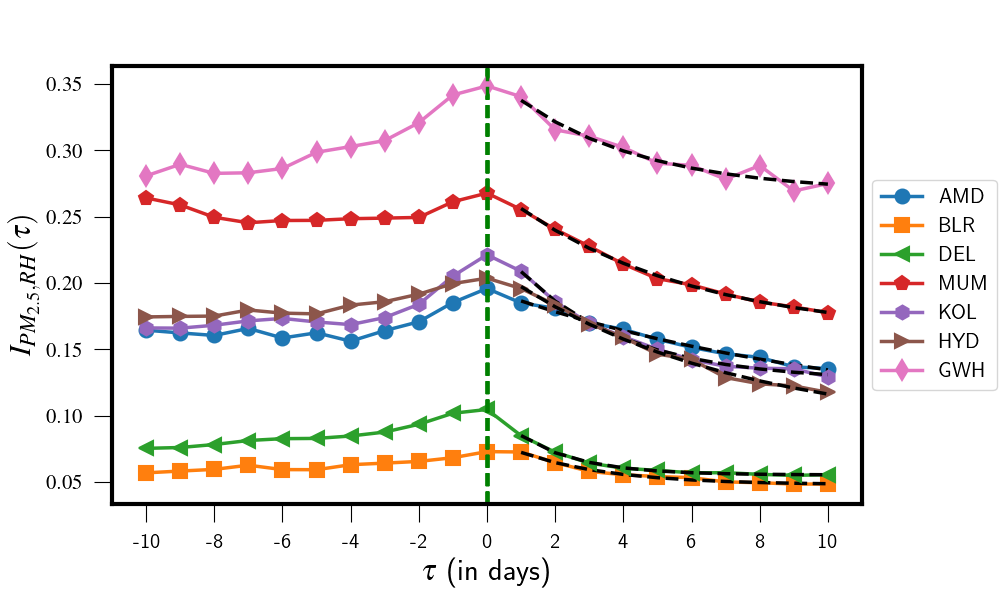}  \includegraphics[width=1.1\linewidth]{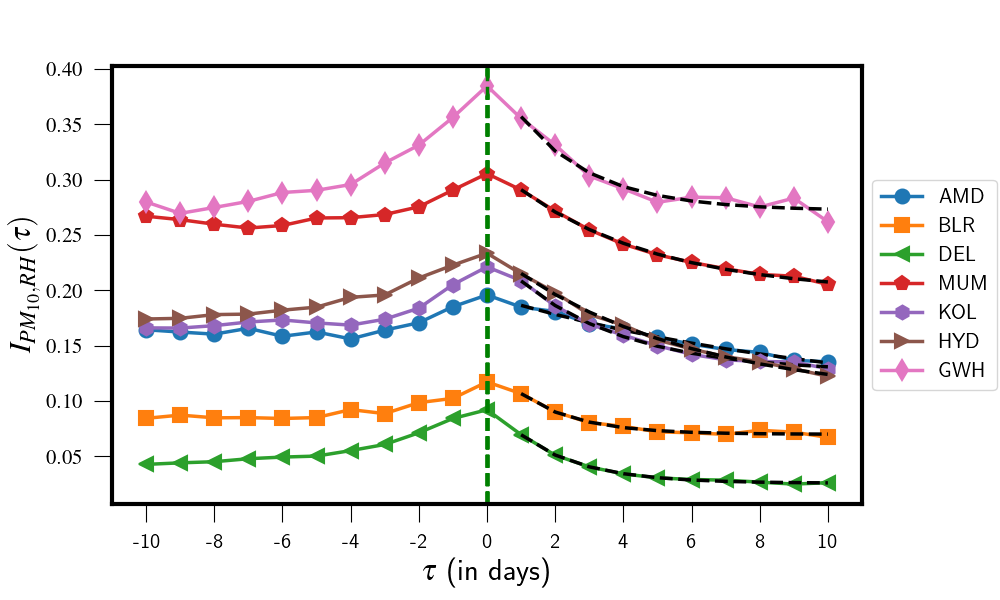}   \includegraphics[width=1.1\linewidth]{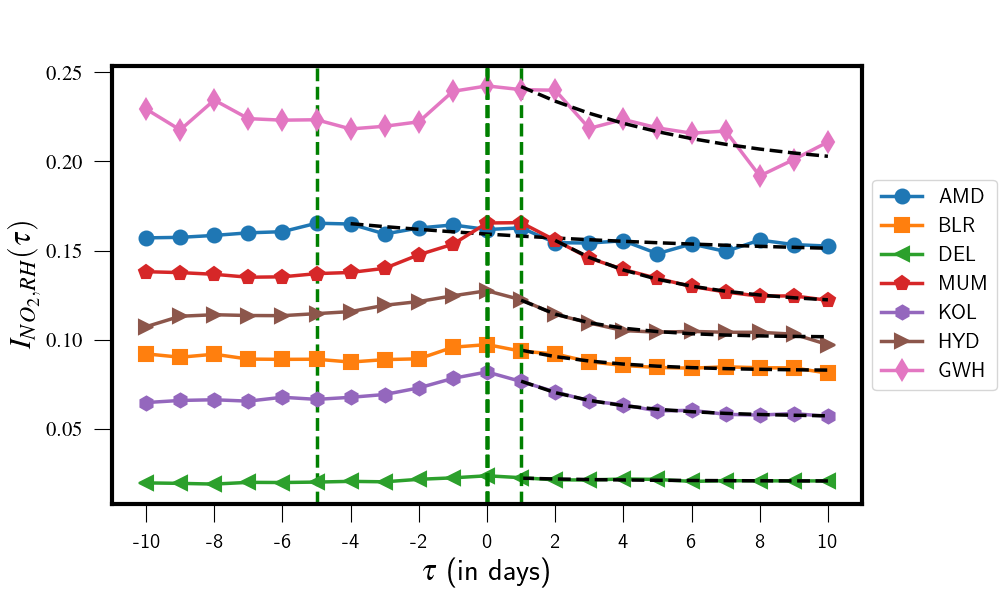}  \includegraphics[width=1.1\linewidth]{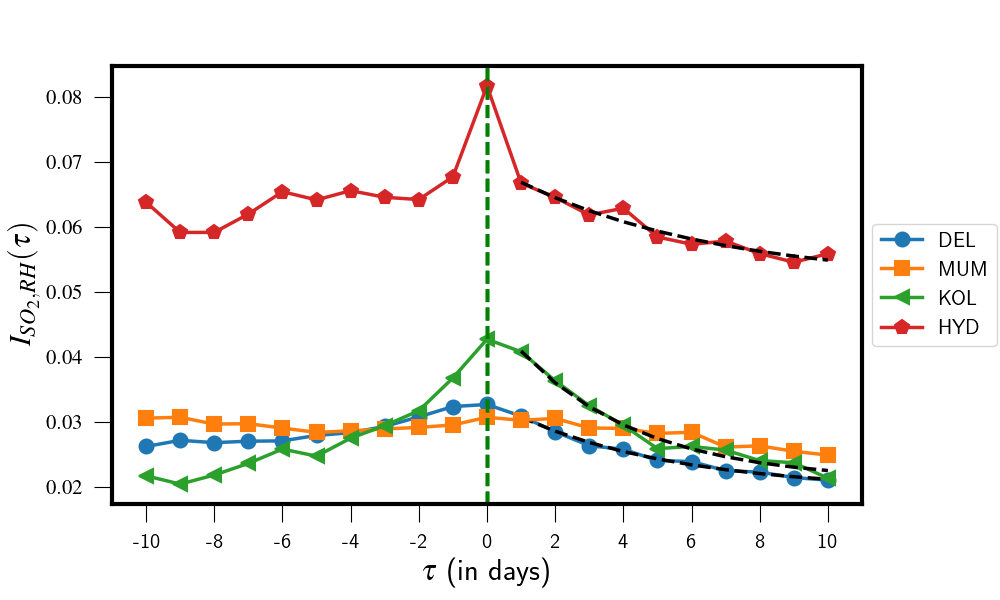}
    \caption{TDMI curves for all four  pollutant-RH pairs across selected representative cities as a function of time-lag $\tau$.}
    \label{fig:TDMI_extra}
\end{figure}

This section presents additional numerical tables and figures containing intermediate results  that are used in the analysis of the main text. These tables and figures are provided for completeness and reproducibility. Fig.~\ref{fig:TDMI_extra} represents the TDMI curves for all four pollutant-RH pairs for some major cities with $\tau\in[-10,10]$.
Table~\ref{tab:1} provides a detailed list of the selected Indian cities. 
Table~\ref{summary} provides a summary showing the city-level mean and standard deviation of each variable across the selected cities, obtained by averaging station-wise statistics. Table~\ref{tab:Entropy} presents the computed self entropy values of all pollutants and meteorological parameters. Further, Table~\ref{tab:Cond_Entropy} gives the conditional entropy values of each pollutant-weather variable pairs and Table~\ref{summary3} provides the summary statistics for the three correlation metrics used in this study. 
Spatial Correlation Score (SCS) values of all the variable pairs are presented in Table~\ref{C_prime}.
\begin{table*}[th]
    \centering
    \small%
    \begin{tabular}{|p{0.4 cm}|p{2.9 cm}|p{2.0 cm}|p{1.5 cm}|c|p{2 cm}|p{1.6 cm}|p{1.1 cm}|}
    \hline
    Sl. No. & City & State/ UT & Population (as of 2024) & $N_m$ & Analysis period & Latitude, Longitude& Climate zone\\ 
   \hline
   \hline 
      1& Ahmedabad (AMD)&Gujarat&90,61,820&3& June 2021 - Dec 2024
& 23.02\textdegree N, 72.57\textdegree E
& BSh
\\
         \hline
      2
& Bengaluru(BLR)&Karnataka&1,43,95,400&5& Jan 2020 - Dec 2024
& 12.97\textdegree N, 77.59\textdegree E
& Aw
\\
         \hline
      3& Chennai (CHN)&Tamil Nadu&1,23,36,000&7& Jan 2022 - Dec 2024& 13.08\textdegree N, 80.27\textdegree E
& As\\
         \hline
      4& Delhi (DEL)&Delhi NCR&3,46,65,600&23& Jan 2020 - Dec 2024
& 28.70\textdegree N, 77.10\textdegree E
& Cwa, BSh\\
         \hline
       5& Hyderabad (HYD)&Andhra Pradesh&1,13,37,900&5& Jan 2020 - Dec 2024& 17.39\textdegree N, 78.49\textdegree E
& BSh\\
         \hline
    6& Kolkata (KOL)&West Bengal&1,58,45,200&7& Jan 2020 - Dec 2024
& 22.57\textdegree N, 88.36\textdegree E
& Aw
\\
         \hline
     7& Mumbai (MUM)&Maharashtra&2,20,89,000 &8 & Jan 2020 - Dec 2024
& 19.08\textdegree N, 72.88\textdegree E
& Aw, Am\\
         \hline
      8& Agartala (AGT)& Tripura & 6,70,388&1& Nov 2020 - Dec 2024& 23.83\textdegree N, 91.29\textdegree E
& Aw\\
         \hline
    9& Asansol (ASN)&West Bengal&15,65,300&1& 
Sept 2022 - Dec 2024
& 23.68\textdegree N, 86.98\textdegree E
& Cwa\\
         \hline
      10& Baddi (BDI)&Himachal Pradesh&29,911&1& Mar 2022 - Dec 2024
& 30.90\textdegree N, 76.80\textdegree E
& Cwa
\\
         \hline
       11& Bhagalpur (BGP)&Bihar&5,25,429&2& 
Jan 2022 - Dec 2024
& 25.25\textdegree N, 86.99\textdegree E
& Cwa
\\
         \hline
       12& Bhopal (BPL)&Madhya Pradesh&26,86,290&3& Feb 2023 - Dec 2024
& 23.26\textdegree N, 77.41\textdegree E
& Cwa
\\
         \hline
          13& Bhubaneswar (BSR)&Odisha&13,20,910&2& Jan 2024 - Dec 2024
& 20.30\textdegree N, 85.82\textdegree E
& Aw
\\
         \hline
      14& Bikaner (BKN)&Rajasthan&8,53,869&1& Feb 2023 - Dec 2024
& 28.02\textdegree N, 73.31\textdegree E
& Bah\\
         \hline
      15& Guwahati (GWH)&Assam&12,24,170&1& Jan 2020 - Dec 2024
& 26.14\textdegree N, 91.74\textdegree E
& Cwa
\\
         \hline
      16& Indore (IND)&Madhya Pradesh&34,82,830&1& Jan 2022 - Dec 2024
& 22.72\textdegree N, 75.86\textdegree E
& Aw\\
         \hline
       17& Kolhapur (KOP)&Maharashtra&6,69,008&2& 
Apr 2023 - Dec 2024
& 16.70\textdegree N, 74.24\textdegree E
& Aw\\
         \hline
       18& Lucknow (LKO)&Uttar Pradesh&41,32,670&4& Jan 2022 - Dec 2024
& 26.85\textdegree N, 80.95\textdegree E
& Cwa
\\
         \hline
          19& Mangaluru (MAN)&Karnataka&7,78,685&1& 
Jul 2021 - Dec 2024
& 12.91\textdegree N, 74.86\textdegree E
& Am
\\
         \hline
    20& Puducherry (PDY)&Puducherry&9,40,911&1& Jan 2022 - Dec 2024
& 11.94\textdegree N, 79.81\textdegree E
& Aw
\\
         \hline
21& Raipur (RPR)&Chhattisgarh&19,23,440&4& Jan 2023 - Dec 2024
& 21.25\textdegree N, 81.63\textdegree E
& Aw
\\
         \hline
    22& Srinagar (SXR)&Jammu and Kashmir&17,77,610&1& 
Jan 2022 - Dec 2024
& 34.08\textdegree N, 74.80\textdegree E
& Cfa\\
         \hline
  23& Thiruvananthapuram (TVM)&Kerala&30,72,530&2& Jan 2020 - Dec 2024
& 8.52\textdegree N, 76.94\textdegree E
&Am
\\\hline
  24& Visakhapatnam (VTZ)&Andhra Pradesh&10,63,178&1& Jan 2020 - Dec 2024
& 17.69\textdegree N, 83.22\textdegree E
&Aw
\\\hline
    \end{tabular}
    \vspace{0.2 cm}
    \caption{List of selected Indian cities and their geographic, climatic, and demographic characteristics. $N_m$ denotes  the number of monitoring stations in each city with a total of 87 stations across all cities.}
    \label{tab:1}
\end{table*}

\begin{table*}
\centering
\small
\begin{tabular}{|l |l |l |l |l |l |l |l |l |l |l |l |l|}\hline
 & \multicolumn{2}{|c|}{PM\textsubscript{2.5}}& \multicolumn{2}{|c|}{PM\textsubscript{10}}& \multicolumn{2}{|c|}{SO\textsubscript{2}}& \multicolumn{2}{|c|}{NO\textsubscript{2}}& \multicolumn{2}{|c|}{RH}& \multicolumn{2}{|c|}{AT}\\\hline\hline             
City & $\langle \rho \rangle$ & $\sigma$ & $\langle \rho \rangle$ & $\sigma$ & $\langle \rho \rangle$ & $\sigma$ & $\langle \rho \rangle$ & $\sigma$ & $\langle \rho \rangle$ & $\sigma$ & $\langle \rho \rangle$ & $\sigma$ \\\hline
AMD & 50.451 & 25.937 & 108.341 & 49.491 & 16.836 & 12.218 & 26.839 & 23.077 & 51.649 & 20.786 & 27.655 & 4.608 \\\hline
BLR & 32.182 & 28.276 & 69.397 & 33.260 & 6.950 & 3.430 & 19.852 & 12.947 & 65.252 & 13.230 & 24.376 & 2.274 \\\hline
CHN & 29.065 & 18.996 & 65.062 & 34.739 & 7.421 & 6.912 & 17.663 & 9.754 & 77.079 & 7.128 & 28.994 & 2.453 \\\hline
DEL & 106.634 & 87.647 & 214.855 & 124.326 & 11.807 & 7.824 & 42.046 & 23.560 & 62.009 & 14.335 & 25.612 & 7.665 \\\hline
HYD & 37.905 & 21.383 & 88.143 & 44.705 & 9.286 & 6.913 & 28.434 & 16.597 & 65.973 & 13.759 & 26.706 & 3.511 \\\hline
KOL & 49.401 & 38.823 & 98.733 & 68.962 & 10.220 & 7.430 & 25.285 & 20.820 & 79.045 & 14.362 & 27.353 & 4.296 \\\hline
MUM & 37.464 & 27.987 & 100.048 & 62.168 & 16.142 & 14.158 & 22.151 & 17.368 & 78.194 & 12.886 & 28.151 & 3.305 \\\hline
AGT & 50.704 & 38.523 & 83.383 & 54.876 & 24.517 & 25.379 & 10.205 & 6.685 & 53.025 & 10.881 & 22.940 & 6.479 \\\hline
ASN & 70.202 & 37.078 & 139.422 & 78.396 & 10.800 & 4.364 & 30.495 & 9.881 & 74.619 & 20.593 & 27.139 & 5.072 \\\hline
BDI & 63.491 & 43.555 & 144.324 & 66.082 & 28.282 & 15.046 & 20.810 & 8.599 & 62.652 & 21.746 & 27.312 & 6.028 \\\hline
BGP & 80.130 & 61.576 & 157.491 & 94.522 & 17.389 & 12.263 & 35.829 & 29.302 & 68.461 & 14.559 & 24.943 & 6.450 \\\hline
BPL & 46.036 & 31.753 & 105.769 & 49.902 & 14.477 & 6.943 & 24.633 & 12.151 & 58.056 & 19.761 & 25.837 & 5.199 \\\hline
BSR & 45.444 & 35.457 & 93.795 & 58.455 & 8.964 & 8.187 & 12.346 & 9.207 & 73.952 & 12.633 & 29.173 & 3.530 \\\hline
BKN & 54.362 & 47.396 & 179.915 & 85.029 & 3.931 & 2.630 & 29.765 & 14.240 & 48.699 & 17.526 & 28.154 & 6.734 \\\hline
GWH & 62.437 & 49.999 & 114.506 & 87.902 & 15.652 & 8.097 & 6.320 & 7.279 & 72.468 & 14.944 & 26.099 & 4.683 \\\hline
IND & 45.585 & 26.748 & 115.525 & 51.205 & 14.307 & 4.921 & 57.984 & 24.716 & 59.077 & 20.443 & 25.770 & 4.572 \\\hline
KOP & 38.671 & 30.063 & 86.741 & 53.292 & 2.420 & 3.786 & 18.993 & 12.605 & 67.963 & 11.120 & 26.888 & 2.032 \\\hline
LKO & 50.752 & 28.472 & 104.108 & 56.409 & 13.183 & 5.510 & 17.068 & 13.737 & 67.841 & 18.040 & 24.718 & 6.670 \\\hline
MAN & 26.157 & 16.072 & 56.569 & 33.164 & 14.575 & 8.730 & 19.663 & 19.324 & 70.974 & 9.834 & 28.700 & 0.820 \\\hline
PDY & 23.886 & 17.123 & 48.549 & 22.422 & 9.529 & 4.291 & 10.543 & 5.066 & 78.152 & 5.990 & 29.501 & 2.371 \\\hline
RPR & 30.792 & 16.936 & 77.046 & 34.212 & 7.826 & 4.492 & 30.220 & 21.095 & 63.094 & 19.594 & 26.633 & 4.540 \\\hline
SXR & 26.619 & 16.496 & 73.758 & 49.008 & 15.906 & 15.432 & 11.691 & 8.869 & 73.028 & 12.429 & 7.939 & 8.690 \\\hline
TVM & 24.445 & 14.112 & 48.432 & 20.567 & 5.874 & 1.786 & 12.876 & 6.880 & 77.380 & 9.244 & 28.186 & 1.532 \\\hline
VTZ & 45.481 & 26.495 & 111.910 & 52.879 & 11.465 & 5.288 & 35.003 & 12.529 & 72.655 & 8.524 & 28.368 & 3.579 \\ \hline

\end{tabular}
\caption{A summary table showing the mean ($\langle \rho \rangle$) and standard deviation ($\sigma$) of each variable, obtained by averaging station-level statistics for all the selected cities.}
\label{summary}
\end{table*}

\begin{table*}[th]
\centering
\small
\begin{minipage}{0.45\textwidth}
\centering
\begin{tabular}{|p{0.7 cm}||c|c|c|c|c|c|}
    \hline
    \textbf{City} & $S_{\text{PM}_{2.5}}$ & $S_{\text{PM}_{10}}$ & $S_{\text{SO}_{2}}$ & $S_{\text{NO}_{2}}$ & $S_{\text{RH}}$ & $S_{\text{AT}}$ \\
    \hline
    AMD & 4.533 & 5.259 & 3.708 & 4.149 & \cellcolor{blue!25}4.394 & 2.941 \\
    \hline
    BLR & 4.167 & 4.888 & 2.477 & 3.801 & 3.987 & 2.205 \\
    \hline
    CHN & 4.204 & 4.849 & 2.854 & 3.696 & 3.531 & 2.350 \\
    \hline
    DEL & \cellcolor{blue!25}6.158 & 5.603 & 3.861 & \cellcolor{blue!25}4.459 & 4.066 & \cellcolor{blue!25}3.338 \\
    \hline
    HYD & 4.336 & 5.120 & 2.932 & 4.215 & 4.039 & 2.695 \\
    \hline
    KOL & 4.745 & 5.368 & 2.951 & 4.083 & 3.925 & 2.799 \\
    \hline
    MUM & 4.465 & 5.399 & 3.629 & 4.016 & 3.860 & 2.636 \\
    \hline
    AGT & 4.631 & 5.091 & 3.835 & 3.128 & 3.784 & 3.163 \\
    \hline
    ASN & 4.866 & 5.607 & 2.739 & 3.591 & 3.927 & 2.994 \\
    \hline
    BDI & 4.809 & 5.509 & \cellcolor{blue!25}3.950 & 3.465 & 4.271 & 3.065 \\
    \hline
    BGP & 5.178 & \cellcolor{blue!25}5.825 & 3.573 & 4.337 & 3.998 & 3.264 \\
    \hline
    BPL & 4.565 & 5.246 & 3.360 & 3.884 & 4.287 & 3.047 \\
    \hline
\end{tabular}
\end{minipage}%
\hfill
\begin{minipage}{0.45\textwidth}
\centering
\begin{tabular}{|p{0.7 cm}||c|c|c|c|c|c|}
    \hline
    \textbf{City} & $S_{\text{PM}_{2.5}}$ & $S_{\text{PM}_{10}}$ & $S_{\text{SO}_{2}}$ & $S_{\text{NO}_{2}}$ & $S_{\text{RH}}$ & $S_{\text{AT}}$ \\
    \hline
    BSR & 4.547 & 5.217 & 3.002 & 3.335 & 3.903 & 2.628 \\
    \hline
    BKN & 4.739 & 5.782 & 2.133 & 3.956 & 4.201 & 3.231 \\
    \hline
    GWH & 4.937 & 5.492 & 3.380 & \cellcolor{red!25}2.677 & 3.918 & 2.873 \\
    \hline
    IND & 4.397 & 5.256 & 2.856 & 4.506 & 4.253 & 2.925 \\
    \hline
    KOP & 4.487 & 5.227 & \cellcolor{red!25}1.776 & 3.762 & 3.622 & 2.109 \\
    \hline
    LKO & 4.680 & 5.378 & 3.258 & 3.658 & 4.207 & 3.285 \\
    \hline
    MAN & 4.021 & 4.739 & 3.241 & 3.745 & \cellcolor{red!25}3.113 & \cellcolor{red!25}0.546 \\
    \hline
    PDY & \cellcolor{red!25}3.911 & \cellcolor{red!25}4.430 & 2.739 & 2.892 & 3.202 & 2.235 \\
    \hline
    RPR & 4.209 & 5.027 & 2.878 & 4.219 & 4.291 & 2.953 \\
    \hline
    SXR & 3.918 & 4.967 & 3.561 & 3.273 & 3.906 & 3.248 \\
    \hline
    TVM & 3.972 & 4.490 & 2.462 & 3.176 & 3.670 & 2.217 \\
    \hline
    VTZ & 4.558 & 5.301 & 2.922 & 3.908 & 3.364 & 2.688 \\
    \hline
\end{tabular}
\end{minipage}
\vspace{0.2 cm}
\caption{Self-entropy values of the selected parameters across cities, highlighting maximum (blue) and minimum (red) values of each variable.}
\label{tab:Entropy}
\end{table*}

\begin{table*}[th]
\centering
\small
 \begin{tabular}{|c||c|c|c|c|c|c|c|c|}
\hline
{\textbf{City}} & ${\cal H}_{\text{PM}_{2.5};\text{RH}}$ & ${\cal H}_{\text{PM}_{2.5};\text{AT}}$ & ${\cal H}_{\text{PM}_{10};\text{RH}}$ & ${\cal H}_{\text{PM}_{10};\text{AT}}$ & ${\cal H}_{\text{SO}_{2};\text{RH}}$ & ${\cal H}_{\text{SO}_{2};\text{AT}}$ & ${\cal H}_{\text{NO}_{2};\text{RH}}$ & ${\cal H}_{\text{NO}_{2};\text{AT}}$\\
\hline
AMD
& 4.364& 4.476& 5.083& 5.204& 3.301& 3.476& 3.696& 3.884
\\\hline
BLR
& 4.122& 4.212& 4.771& 4.847& 2.411& 2.471& 3.702&3.765
\\\hline
 CHN
& 4.001& 4.033& 4.706& 4.713& 2.805& 2.811& 3.522&3.582
\\\hline
DEL
& \cellcolor{blue!25}5.464& \cellcolor{blue!25}5.335& \cellcolor{blue!25}6.052& \cellcolor{blue!25}6.016& 3.240& 3.283& \cellcolor{blue!25}4.502& \cellcolor{blue!25}4.452
\\
    \hline
    HYD
& 4.142& 4.264& 4.890& 5.045& 2.867& 2.836& 4.096& 4.119
\\
    \hline
KOL
& 4.610& 4.336& 5.213& 4.948& 2.936& 2.784& 4.054& 3.848
\\
    \hline
MUM
& 4.278& 4.393& 5.160& 5.347& \cellcolor{blue!25}3.639& 3.585& 3.915& 3.960
\\\hline
 AGT
& 4.244& 4.265& 4.796& 4.760& 3.620& 3.602& 2.960&2.945
\\\hline
 ASN
& 4.158& 4.524& 4.812& 5.233& 2.297& 2.604& 3.045&3.289
\\
    \hline
BDI
& 4.339& 4.449& 5.158& 5.356& 3.612& \cellcolor{blue!25}3.762& 3.269& 3.216
\\
    \hline
BGP
&  4.836&  4.713&  5.595&  5.601&  3.459&  3.474&  3.923& 3.962
\\
    \hline  
BPL
& 4.259& 4.224& 4.950& 5.044& 3.273& 3.321& 3.650& 3.747
\\
    \hline    
BSR
&  4.173&  3.917&  4.784&  4.637&  2.738&  2.677&  2.825& 2.555
\\
    \hline
BKN
&  4.248&  4.235&  5.448&  5.621&  1.917&  1.970&  3.724& 3.803
\\
    \hline
GWH
&  4.559&  4.454&  5.124&  5.058&  2.883&  2.946&  \cellcolor{red!25}1.942& \cellcolor{red!25}1.728
\\
    \hline
IND
&  4.007&  4.205&  4.801&  5.128&  2.56&  2.830&  4.059& 4.407
\\
    \hline  
KOP
&  3.945&  4.355&  4.656&  5.112&  \cellcolor{red!25}1.025&  \cellcolor{red!25}1.471&  3.284& 3.660
\\
    \hline
    LKO
& 4.468& 4.527& 5.106& 5.246& 3.121& 3.215& 3.339& 3.473
\\
    \hline
 MAN
& 3.909& 3.980& 4.641& 4.673& 3.040& 3.108& 3.313&3.504
\\\hline
 PDY
& \cellcolor{red!25}3.771& 3.731& \cellcolor{red!25}4.319& 4.359& 2.702& 2.709& 2.852&2.707
\\\hline
 RPR
& 3.930& 4.082& 4.756& 4.929& 2.618& 2.702& 3.741&3.978
\\ \hline
SXR
& 3.860& 3.795& 4.871& 4.792& 3.406& 3.382& 3.104&3.021
\\\hline
 TVM
& 3.797& \cellcolor{red!25}3.695& 4.347& \cellcolor{red!25}4.217& 2.161& 2.003& 3.052&2.994
\\\hline
VTZ
& 4.305& 4.191& 5.056& 4.981& 2.700& 2.683& 3.710&3.637
\\\hline
    \end{tabular}
    \vspace{0.2 cm}
\caption{City-wise conditional entropy values of air pollutants with respect to meteorological parameters highlighting maximum (blue) and minimum (red) values.}
    \label{tab:Cond_Entropy}
\end{table*}

\begin{table*}
\centering
\small
\begin{tabular}{|
>{\raggedright\arraybackslash}p{2cm}|
>{\raggedright\arraybackslash}p{1.5cm}|
>{\raggedright\arraybackslash}p{1cm}|
>{\raggedright\arraybackslash}p{1cm}|
>{\raggedright\arraybackslash}p{1cm}|
>{\raggedright\arraybackslash}p{1cm}|
>{\raggedright\arraybackslash}p{0.9cm}|
>{\raggedright\arraybackslash}p{0.5cm}|}
\hline
pollutant--weather variable pairs
& \multicolumn{3}{c|}{Pearson correlation ($r_{X,Y}$)}
& \multicolumn{2}{c|}{\shortstack{Mutual\\ information ($I_{X,Y}$)}}
& \multicolumn{2}{c|}{\shortstack{Relative \\conditional\\entropy ($\mathcal{H}^R_{X,Y}$)}} \\  
\hline
& $\%$ of cities with $r_{X,Y}<0$ & Median & $\sigma$
& Median & $\sigma$
& Median & $\sigma$ \\
\hline
PM\textsubscript{2.5}-RH & 75.00 & -0.130 & 0.182 & 0.249 & 0.137 & 0.939 & 0.034 \\
\hline
PM\textsubscript{2.5}-AT & 79.17 & -0.300 & 0.243 & 0.247 & 0.181 & 0.942 & 0.039 \\
\hline
PM\textsubscript{10}-RH & 95.83 & -0.285 & 0.207 & 0.301 & 0.163 & 0.955 & 0.039 \\
\hline
PM\textsubscript{10}-AT & 87.50 & -0.210 & 0.234 & 0.281 & 0.164 & 0.972 & 0.035 \\
\hline
SO\textsubscript{2}-RH & 83.33 & -0.140 & 0.217 & 0.213 & 0.147 & 0.931 & 0.085 \\
\hline
SO\textsubscript{2}-AT & 87.50 & 0.000 & 0.215 & 0.190 & 0.168 & 0.951 & 0.052 \\
\hline
NO\textsubscript{2}-RH & 66.67 & -0.195 & 0.212 & 0.229 & 0.156 & 0.942 & 0.060 \\
\hline
NO\textsubscript{2}-AT & 58.33 & -0.180 & 0.226 & 0.268 & 0.183 & 0.943 & 0.074 \\
\hline
\end{tabular}
\vspace{0.2 cm}
\caption{Summary statistics of the correlation metrics – Pearson correlation, mutual information, and relative conditional entropy for pollutant–meteorological variable pairs across all cities.}
\label{summary3}
\end{table*}

\begin{table*}[th]
\centering
\small
\begin{tabular}{|l |l |l |l |l |l |l |l |l|}\hline 
{\textbf{City}} & $\phi$\textsubscript{{\tiny PM}\textsubscript{2.5},\text{\tiny RH}} & $\phi$\textsubscript{{\tiny PM}\textsubscript{2.5},\text{\tiny AT}} & $\phi$\textsubscript{{\tiny PM}\textsubscript{10},\text{\tiny RH}} & $\phi$\textsubscript{{\tiny PM}\textsubscript{10},\text{\tiny AT}} & $\phi$\textsubscript{{\tiny SO}\textsubscript{2},\text{\tiny RH}} & $\phi$\textsubscript{{\tiny SO}\textsubscript{2},\text{\tiny AT}} & $\phi$\textsubscript{{\tiny NO}\textsubscript{2},\text{\tiny RH}} & $\phi$\textsubscript{{\tiny NO}\textsubscript{2},\text{\tiny AT}}\\\hline
AMD & 1.18& 1.09& 2.83& 0.55& 1.18& 0.58& 1.89& 0.58
\\\hline
BLR & 0&0& 2.40& 0.05& 0.41& 0& 1.00&0
\\\hline
CHN & 0.58& 1.77& 1.60& 0.95& 0.43& 1.27& 0.61& 0.61
\\\hline
DEL & 2.23& 4.37& 0& 0& 2.40& 0.38& 0& 1.04
\\\hline
HYD & 1.61& 0.96& 3.71& 0.04& 0.80& 1.44& 1.41& 0.47
\\\hline
KOL & 1.35& 3.59& 3.34& 3.33& 1.61& 1.05& 1.04& 1.32
\\\hline
MUM & 2.07& 1.05& 4.14& 0.28& 0& 0.16& 2.08& 0.56
\\\hline
AGT & 3.03& 3.47& 3.03& 3.42& 1.82& 3.45& 2.04& 2.24
\\\hline
ASN & \cellcolor{blue!25}5.60& 4.30& 6.31& 4.38& 4.93& 3.84& 4.37& 3.64
\\\hline
BDI & 2.94& 4.36& 3.38& 2.69& 2.24& \cellcolor{blue!25}4.47& 2.13& 4.00
\\\hline
BGP & 1.97& 4.43& 3.04& 2.35& 1.01& 0.80& 2.24& 2.10
\\\hline
BPL & 2.46& 3.82& 4.14& 2.47& 1.69& 0.95& 3.48& 2.26
\\\hline
BSR & 3.73& \cellcolor{blue!25}6.22& 4.82& \cellcolor{blue!25}5.53& 2.59& 2.48& 3.89& \cellcolor{blue!25}6.30\\\hline
BKN & 3.95& 4.87& 5.42& 3.00& 2.78& 3.50& 3.47& 2.75
\\\hline
GWH & 2.60& 4.30& 3.56& 3.81& 3.78& 2.74& 3.30& 4.11
\\\hline
IND & 2.46& 2.26& 5.22& 1.20& 3.33& 1.23& 3.91& 1.41
\\\hline
KOP & 4.64& 1.09& \cellcolor{blue!25}6.51& 1.06& \cellcolor{blue!25}5.18& 1.10& \cellcolor{blue!25}4.97& 0.82
\\\hline
LKO & 0.93& 2.16& 3.13& 1.25& 1.25& 0.80& 1.62& 0.87
\\\hline
MAN & 1.43& 0.86& 2.69& 0.68& 3.56& 2.97& 3.40& 2.12
\\\hline
PDY & 0.98& 2.25& 3.07& 0.81& 1.83& 2.15& 0.87& 2.57
\\\hline
RPR & 2.27& 1.78& 3.93& 1.31& 2.73& 2.53& 3.68& 1.12
\\\hline
SXR & 0.73& 1.33& 1.98& 1.23& 1.10& 1.74& 1.69& 2.49
\\\hline
TVM & 1.40& 2.19& 2.79& 1.98& 1.66& 2.89& 0.66& 0.81
\\\hline
VTZ & 1.52& 3.35& 2.66& 2.00& 1.54& 0.93& 1.39& 1.15
\\
\hline
\end{tabular}
\vspace{0.2 cm}
\caption{SCS ($\phi$) values for 8 pollutant-weather variable pairs across 24 cities. 
The highest $\phi$ values for each interaction pair are highlighted with blue color.}
\label{C_prime}
\end{table*}
\bibliography{ref}

\end{document}